\documentclass[twocolumn,superscriptaddress]{aastex7}

\usepackage[percent]{overpic}
\usepackage{graphicx}
\usepackage{subfigure}
\usepackage{subcaption}
\usepackage{placeins}
\usepackage{bm}
\usepackage{xcolor}
\usepackage{float}
\usepackage{amsmath}
\usepackage{ulem}

\usepackage[inline]{enumitem}

\shorttitle{ 
Einstein@Home search for Continuous Gravitational Waves from Cas A,  Vela Jr. and G347.3
}
\shortauthors{Ming et al.}

\begin{document}


\title{ 
Observational constraints on the spin/anisotropy of the CCOs of Cassiopeia A, Vela Jr. and G347.3-0.5 and a single surviving continuous gravitational wave candidate 
}

\author{Jing Ming}
\email{jing.ming@aei.mpg.de}
\affiliation{ Max Planck Institute for Gravitational Physics (Albert Einstein Institute), Callinstrasse 38, 30167, Hannover, Germany }
\affiliation{ Leibniz Universit{\"a}t Hannover, D-30167 Hannover, Germany }

\author{Maria Alessandra Papa}
\email{maria.alessandra.papa@aei.mpg.de}
\affiliation{ Max Planck Institute for Gravitational Physics (Albert Einstein Institute), Callinstrasse 38, 30167, Hannover, Germany }
\affiliation{ Leibniz Universit{\"a}t Hannover, D-30167 Hannover, Germany }

\author{Heinz-Bernd Eggenstein}
\email{heinz-bernd.eggenstein@aei.mpg.de}
\affiliation{ Max Planck Institute for Gravitational Physics (Albert Einstein Institute), Callinstrasse 38, 30167, Hannover, Germany }
\affiliation{ Leibniz Universit{\"a}t Hannover, D-30167 Hannover, Germany }

\author{Bernd Machenschalk}
\email{Bernd.Machenschalk@aei.mpg.de}
\affiliation{ Max Planck Institute for Gravitational Physics (Albert Einstein Institute), Callinstrasse 38, 30167, Hannover, Germany }
\affiliation{ Leibniz Universit{\"a}t Hannover, D-30167 Hannover, Germany }

\author{J. Martins}
\email{Jasper.Martins@aei.mpg.de}
\affiliation{ Max Planck Institute for Gravitational Physics (Albert Einstein Institute), Callinstrasse 38, 30167, Hannover, Germany }
\affiliation{ Leibniz Universit{\"a}t Hannover, D-30167 Hannover, Germany }

\author{B. Steltner}
\email{benjamin.Steltner@aei.mpg.de}
\affiliation{ Max Planck Institute for Gravitational Physics (Albert Einstein Institute), Callinstrasse 38, 30167, Hannover, Germany }
\affiliation{ Leibniz Universit{\"a}t Hannover, D-30167 Hannover, Germany }

\author{B. McGloughlin}
\email{Brian.McGloughlin@aei.mpg.de}
\affiliation{ Max Planck Institute for Gravitational Physics (Albert Einstein Institute), Callinstrasse 38, 30167, Hannover, Germany }
\affiliation{ Leibniz Universit{\"a}t Hannover, D-30167 Hannover, Germany }

\author{V. Dergachev}
\email{Vladimir.Dergachev@aei.mpg.de}
\affiliation{ Max Planck Institute for Gravitational Physics (Albert Einstein Institute), Callinstrasse 38, 30167, Hannover, Germany }
\affiliation{ Leibniz Universit{\"a}t Hannover, D-30167 Hannover, Germany }

\author{R. Prix}
\email{Reinhard.Prix@aei.mpg.de}
\affiliation{ Max Planck Institute for Gravitational Physics (Albert Einstein Institute), Callinstrasse 38, 30167, Hannover, Germany }
\affiliation{ Leibniz Universit{\"a}t Hannover, D-30167 Hannover, Germany }

\author{M. Bensch}
\email{maximillian.bensch@aei.mpg.de}
\affiliation{ Max Planck Institute for Gravitational Physics (Albert Einstein Institute), Callinstrasse 38, 30167, Hannover, Germany }
\affiliation{ Leibniz Universit{\"a}t Hannover, D-30167 Hannover, Germany }

\correspondingauthor{J. Ming}
\email{jing.ming@aei.mpg.de}
\correspondingauthor{M.Alessandra Papa}
\email{maria.alessandra.papa@aei.mpg.de}

\begin{abstract}

We carry out the deepest and broadest search for continuous gravitational-wave signals from three neutron stars at the center of the supernova remnants Cassiopeia A, Vela Jr., and G347.3-0.5. This search was made possible by the computing power shared by thousands of Einstein@Home volunteers. After the initial Einstein@Home search, which used O3a data, we perform a multi-stage follow-up of the most promising $\approx$ 45 million signal candidates. In the last stages, we use independent data (O3b and O4a) to further investigate the remaining candidates from the previous stages. We set the most stringent constraints to date on the gravitational-wave amplitude, equatorial ellipticity, r-mode saturation amplitude, and -- for the first time -- the neutron-star crustal anisotropy.  For spin periods lower than 2 ms we constrain the ellipticity to be {\it{smaller}} than $4\times 10^{-7}$ for all targets. We exclude crustal anisotropy values larger than $5\times 10^{-3}$ for spin periods between 1.3--100 ms. Only one candidate -- from the G347.3 search -- survives all follow-ups. We illustrate properties of this candidate. Investigations on new data will aid in clarifying its nature. Such ``new" data  already exist, O4b and O4c, and would be optimal for this purpose, but they are not publicly accessible at the time of writing. In the appendix we provide our estimate of the candidate phase parameters, which are useful for others to carry out checks on the new data.

\end{abstract}



\section{\label{sec:intro}Introduction}

Continuous gravitational waves (CWs) are long-lived, nearly monochromatic
signals whose amplitudes are expected to be orders of magnitude below the
bursts produced by coalescing compact binaries \cite[see][for burst amplitudes] {LIGOScientific:2025slb} and \cite[][and references therein, for expected CW amplitudes]{Jones:2024nty}. 

Ten years after the first detection of gravitational waves \citep{GW150914}, continuous waves have not yet been detected.
 To increase the chances of detection, one must integrate the detector strain data coherently over the longest possible observation spans (months), thereby accumulating signal power with time and boosting the signal-to-noise ratio roughly as $\propto T_\mathrm{coh}^{1/2}$.
 In practice, however, extending the coherent time $T_\mathrm{coh}$ enlarges the size of the template bank dramatically and this makes coherent searches over months impossible over large parameter spaces \citep{Brady:1997ji,Krishnan:2004sv,Riles:2022wwz}.

Among many astrophysical sources of continuous gravitational waves, supernova remnants are particularly interesting because the embedded neutron stars are young enough to support relatively large non-axisymmetric deformations \cite[as first proposed by][]{LIGOScientific:2008hqb} or unstable $r$-modes that could power continuous gravitational-wave emission higher than for older neutron stars \citep{Andersson1999, Ushomirsky2000, Arras_2003, Haskell2006, Gittins2023}. Moreover, their sky positions are known, and their distances and ages are often constrained by multi-wavelength observations, making the search easier: fixing (or tightly constraining) the sky location collapses the search parameter space and permits longer coherent integration than all-sky surveys, e.g. \cite{Abbott2022_3,Steltner2023,Dergachev2023,covas2024,brian2025a,brian2025b}. 
Consequently, directed searches toward supernova remnants can reach substantially greater depth than contemporary wide-parameter all-sky surveys at comparable computational cost.
Moreover, age ($\tau$) and distance ($D$) estimates provide informative priors on the spin-down range and yield the so-called age-based indirect gravitational wave amplitude upper limit \citep{LIGOScientific:2008hqb}:
\begin{equation}
h_0^{\text{age}} \leq 1.22 \times 10^{-24}
\left( \frac{3.4 \,\text{kpc}}{D} \right)
\sqrt{
\left( \frac{300 \,\text{years}}{\tau} \right)
}
\label{eq:h0_age}
\end{equation}
which corresponds to the optimistic case in which the star’s rotational energy loss over its lifetime has been dominated by gravitational radiation. 

In our galaxy, out of the $310$ Galactic supernova remnants cataloged by \cite{Green:2024uci,Green_2024}, fifteen young systems have been prioritized as plausible continuous-wave targets \citep{LIGOScientific:2021mwx}.
In \cite{Ming2016}, we introduced an optimization framework to allocate computational power efficiently across searches for different supernova remnants. Applying that scheme identifies three remnants as the highest-return investments for deep searches: Cassiopeia A (G111.7$-$2.1), Vela Jr.\ (G266.2$-$1.2),  and G347.3 (G347.3$-$0.5), hereafter referred to as Cas A, Vela Jr. and G347.3, respectively.

Since the completion of the third LIGO observing run (O3) in 2020, several searches targeting these three sources have been performed \citep{LIGOScientific:2021mwx,Abbott2022_4,wang2024,Salvadore2025}. Our Einstein@Home group has previously published results for these targets using data from the first (O1) and second (O2) LIGO observing runs \citep{Ming2019,Papa_2020midth,Ming2022,ming2024a,ming2025}. In this paper we present, for the first time, results for all three targets based on O3 data.

This manuscript is organized as follows. 
Section~\ref{sec:targets} introduces the targeted remnants; 
Section~\ref{sec:signal_model}  describes the signal model for continuous gravitational waves;
Section~\ref{sec:data} describes the data sets that we use and their cleaning/conditioning; 
Section~\ref{sec:search} describes the Einstein@Home search pipeline;
Section~\ref{sec:FU} describes the hierarchical follow-up searches and their results;
finally, Section~\ref{sec:results} reports astrophysical constraints and discusses the astrophysical implications.


\section{\label{sec:targets} Targets}


\subsection{\label{subsec:cas_a} Cassiopeia A (G111.7-2.1)}

Cassiopeia A (CasA) hosts a central compact object (CCO) formed in one of the most recent Galactic core-collapse supernovae. The CCO, CXOU J232327.9+584842, lies at the center of the Cas~A supernova remnant, and its position was measured with the \textit{Chandra} X-ray satellite \citep{Tanabaum1999}. We use $(\alpha,\delta) = \left(23^{\mathrm h}23^{\mathrm m}27.9^{\mathrm s},\, +58^\circ48'42.4''\right)$ as the position of the neutron star. The prevailing consensus dates the explosion to $\sim310$--$350$~yr ago \citep{Fesen2006} at a distance of $3.3$--$3.7$~kpc from Earth \citep{Reed1995}, based on the observed expansion of the outer ejecta and the radial motion of the CCO. From its X-ray spectrum, \cite{Ho2009} suggest that the central object is a neutron star with a carbon atmosphere and a relatively small magnetic field. The intricate, asymmetric morphology of the surrounding remnant likely reflects a non-spherical explosion which may have produced a non-axisymmetric neutron star. Additionally, if the newly born neutron star was rapidly spinning and indeed had a relatively weak internal magnetic field, as \cite{Ho2009} suggest, it could have been susceptible to rotational instabilities such as r-modes \citep{Owen:1998xg}, potentially powering long-lived, nearly monochromatic gravitational-wave emission.
In this work, we take 3.4 kpc to be the distance and 330 years to be the age of the Cas A.


\subsection{\label{subsec:vela_jr} Vela Jr. (G266.2-1.2)}

The CCO associated with the Vela Jr.\ (G266.2$-$1.2) supernova remnant is CXOU J085201.4$-$461753. A large X-ray-to-optical flux ratio, together with the absence of a bright optical counterpart, is consistent with the CCO being an isolated neutron-star \citep{Pavlov2001}. The source position was first pinpointed with \textit{Chandra} \citep{Pavlov2001}, and subsequent near-infrared observations refined the astrometry and further supported the neutron-star hypothesis \citep{Mignani2007}. We use $(\alpha,\delta) = \left(08^{\mathrm h}52^{\mathrm m}01.4^{\mathrm s},\, -46^\circ17'53.3''\right)$ as the position of the neutron star.

Inferences from Ti$^{44}$ line emission favour a very young and nearby remnant, with an age of $\sim700$~yr at a distance of $\sim200$~pc \citep{Iyudin1998}. In contrast, analyses based on \textit{Chandra} X-ray data combined with hydrodynamic modelling of the remnant’s expansion point to an older and more distant object, $\sim4300$~yr at $\sim750$~pc \citep{Allen2015}. In what follows we treat these two $(\tau,D)$ solutions as bracketing scenarios: $(\tau=700~\mathrm{yr}, D=200~\mathrm{pc})$ to be optimistic and $(\tau=4300~\mathrm{yr}, D=750~\mathrm{pc})$ to be pessimistic.


\subsection{\label{subsec:3473} G347.3-0.5}

The supernova remnant G347.3 has been proposed as the remnant of the AD~393 ``guest star'' \citep{1997A&A...318L..59W}. Adopting this association yields an age of $\sim1600$~yr, though the identification is not completely uncontroversial  \citep{2012AJ....143...27F}. The most recent distance estimate at $1.12 \pm 0.01~\mathrm{kpc}$  by \cite{Leike:2020jyl} holds a very tight statistical uncertainty, but the actual uncertainty may be larger due to model assumptions. XMM observations place the remnant at a distance of $\sim1.3$ kpc \citep{2004A&A...427..199C}. Conservatively we adopt this distance when translating the gravitational wave amplitude upper limits into physical constraints on the neutron star, as this larger distance produces less stringent constraints. The position of its central compact object has been measured with sub-arcsecond accuracy using \textit{Chandra} data \citep{2008A&A...484..457M}. We use $(\alpha,\delta) = \left(17^{\mathrm h}13^{\mathrm m}28.3^{\mathrm s},\, -39^\circ49'53.3''\right)$ as the position of the neutron star.


\section{\label{sec:signal_model} The Signal Model}

A gravitational wave is described in the plane perpendicular to the direction of propagation by the two strains $h_{+}$ and $h_{\times}$, corresponding to the two polarizations:
\begin{equation}
h_+ (t) = A_+ \cos \Phi (t),
\end{equation}
\begin{equation}
h_{\times} (t) = A_{\times} \sin \Phi (t).
\end{equation}
$\Phi (t)$ is the gravitational-wave phase and $A_{+,\times}$ denote the gravitational-wave polarization amplitudes:
\begin{equation}
A_{+} = \frac{1}{2} h_0 (1 + \cos^2 \iota),
\end{equation}
\begin{equation}
A_{\times} = h_0 \cos \iota, 
\end{equation}
where $\iota$ represents the angle between the angular momentum of the neutron star and the line of sight from Earth, while $h_0$ is the intrinsic wave amplitude. 

The frequency of a continuous wave emitted by an isolated rapidly rotating neutron star evolves gradually over time, and its time behaviour can be described using a Taylor expansion around the values at a reference time $\tau_0$ in the Solar System Barycenter (SSB):
\begin{equation}
f(\tau) = f_0 + \dot{f}_0 (\tau - \tau_0) + \frac{1}{2} \ddot{f}_0 (\tau - \tau_0)^2.
\end{equation}
We adopt SSB time  $\tau_0=1246197626.5$ s (GPS) for all the searches using O3 data and  $\tau_0=1379214442$ s  (GPS) for the searches using O4a data. These reference times approximately correspond to the mid time of O3a and O4a, respectively. This is a standard choice for continuous wave searches that minimizes the amount of input data needed for the search and maximizes computational efficiency  \cite[see, for example ][that uses O3a data and whose reference time differs from ours by less than 1.5 days]{Dergachev:2024knd}.

The relationship between the SSB time $\tau$ and the detector time $t$ is defined by the condition that
\begin{equation}
\Phi(\tau)=\Phi^\prime(\tau(t)),
\label{eq:tau_bary}
\end{equation}
where $\Phi^\prime(t)$ is the signal phase measured at detector time $t$. For an isolated source, the $\tau(t)$ transformation encapsulates the R{\text{\o}}mer, Einstein and Shapiro delays.

In the detector data, the continuous wave signal has the form \citep{JKS1998}:
\begin{equation}
h(t)=F_+(t)\,h_+(t)+F_{\times}(t)\,h_{\times}(t),
\label{eq:signal}
\end{equation}
where $F_+(t)$ and $F_\times(t)$ are the detector beam-pattern functions which depend on the source sky position $(\alpha,\delta)$ and on the angle $\psi$ of the wave frame with respect to the detector frame. Because the detector rotates with the Earth, $F_{+,\times}(t)$ exhibit sidereal-day periodicity.


\section{\label{sec:data} The Data}

For the Einstein@Home searches we use publicly available data from the first half of the third LIGO observing run (O3a), spanning GPS times 1238421231 s (April 04 2019) to 1253973231 s  (October 01 2019) \citep{Abbott_2023}. 
For the post-processing of the Einstein@Home outliers we additionally use data from the second half of the third observing run (O3b) and the first part of the fourth observing run (O4a).

We use the {\texttt{GWOSC-16KHZ\_R1\_STRAIN}} channel for O3 and the {\texttt{GDS-CALIB\_STRAIN\_NOLINES\_AR}} channel for O4a, with the CBC CAT2 data quality flags for all runs. With these choices the O3a Hanford detector (LHO) has a duty factor of  $71\%$ whereas the Livingston detector (LLO) a duty factor of $76\%$. O3b data spans GPS times 1256655667 s (November 01 2019) to 1269361693 s (March 27 2020), with slightly higher duty factors of $79\%$ for both detectors than O3a \citep{Abbott_2023,2023PhRvX..13d1039A}. O4a data spans GPS times 1368975618 s (May 24 2023) to 1389456018 s (January 16 2024), with slightly lower duty factors of $69\%$ for both LLO and $68\%$ for LHO than O3a \citep{LIGOScientific:2025pvj,DiCesare:2025wnb}.

Short Fourier Transforms of 1800-second data segments are 
used as the search input. Prominent instrumental features—calibration lines, mains-power harmonics, and spurious noise associated with laser-beam jitter—were removed from the data before it was released \citep{Davis_2019,Vajente_2020}. In addition, we excise loud, short-duration glitches and further   
replace frequency bins contaminated by lines \citep{o3_linefile,o4_linefile} with Gaussian noise consistent with the local spectrum. Our data-preparation is described in \cite{Steltner2022}.


\section{\label{sec:search} The Einstein@Home Search}

Our analysis adopts a ``stack–slide'' type \citep{Brady:1997ji,2000PhRvD..61h2001B} semi-coherent search implemented with the Global Correlation Transform (GCT) method \citep{Pletsch2008,Pletsch2009,Pletsch2010}. We divide the total observing span $T_{\text{obs}}$ into $N_{\text{seg}}$ equal-length segments of coherent duration $T_{\text{coh}}$. For each segment $i$, using the data from both interferometers, we compute the maximum-likelihood coherent multi-detector ${\mathcal{F}}$-statistic \citep{Cutler:2005hc} by matched filtering the data against a phase-evolution model with parameters ${f,\dot f,\ddot f,\alpha,\delta}$, while analytically maximizing over the amplitude parameters ${h_0,\iota,\psi,\Phi_0}$. The semi-coherent detection statistic is 
\begin{equation}
   \Bar{\mathcal{F}} = \frac{1}{N_\mathrm{seg}} \sum_{i=1}^{N_\mathrm{seg}} \mathcal{F}_i\,,
   \label{eqn:mean_2F}
\end{equation}
with each $\mathcal{F}_i$ coming from a suitable point in parameter space, that depends on the phase model parameters for which we want to compute $\Bar{\mathcal{F}}$.

In stationary Gaussian noise,  the semi‑coherent sum $N_{\mathrm{seg}} \times 2\overline{\mathcal{F}}$ is $\chi^2$–distributed with $4N_{\mathrm{seg}}$ degrees of freedom. 
In the presence of a signal, the same quantity follows a non‑central $\chi^2$–distribution $\chi^2_{4N_{\mathrm{seg}}}(\rho^2)$, where the non-centrality parameter
\begin{equation}
\label{eq:rho2}
\rho^2\propto {h_0^2 T_{\text{obs}} \over {S_h}}.
\end{equation}
Here $S_h(f)$ denotes the one-sided spectral density of the detector’s noise, evaluated at the signal’s frequency \citep{JKS1998}.

Although many prominent lines in the data we use have been removed, some weaker coherent-features can survive and bias the semi‑coherent statistic $\Bar{\mathcal{F}}$. To mitigate these high $\Bar{\mathcal{F}}$, we rank Einstein@Home candidates with the \emph{line‑robust} statistic $\hat{\beta}_{\mathrm{S/GLtL}}$, defined as the logarithm of the Bayesian odds in favor of a continuous‑wave signal (S) versus an extended noise model that includes stationary Gaussian noise (G), persistent single‑detector lines (L), and short‑lived line transients (tL) \citep{Keitel:2013wga,Keitel:2015ova}. Using $\hat{\beta}_{\mathrm{S/GLtL}}$ in place of a pure  $\Bar{\mathcal{F}}$ ranking substantially reduces line‑like outliers that resemble signals.

The per‑segment coherent statistic $2\mathcal{F}_i$ combines data from both detectors within segment $i$, whereas the cross‑segment accumulation in Eq.~(\ref{eqn:mean_2F}) is incoherent. This mixed strategy is the reason the overall procedure is termed \emph{semi‑coherent}. For computational efficiency, the initial evaluation of the semi-coherent detection statistics employs the GCT approximation at each template. Candidates that rise to the top lists are then \emph{recomputed} exactly at their nominal parameter points; we denote these refined values with the subscript ``r'', e.g  $2{\mathcal{F}_r}$ and $\hat{\beta}_{\mathrm{S/GLtL}r}$. True signals typically increase under this exact recomputation, while many noise outliers do not.

We adopt a hierarchical scheme consisting of multiple semi‑coherent stages. Sensitivity increases with stage number, so a signal is expected to gain significance as it progresses, whereas random or instrumental artifacts generally fail to do so. The Einstein@Home search, which is  also denoted as Stage-0 search, covers the full parameter space and therefore dominates the total computational cost of the search. Subsequent stages restrict the follow‑up search in the parameter volume around each surviving candidate and apply more sensitive search set-ups; only candidates consistent with signal‑like behavior are promoted to the next stage. In total we use five stages (0–4), progressively pruning the top list until only the most significant candidates survive.

Key configuration parameters for any stage are: the segment coherence time $T_\mathrm{coh}$, the number of segments $N_\mathrm{seg}$, the total timespan $T_\mathrm{obs}$, the template‑bank spacings in frequency and spindown $\{\delta f,\delta\dot f,\delta\ddot f\, , \gamma_1,  \gamma_2\}$.

$\{\delta f,\delta\dot f,\delta\ddot f\}$  define the grid used for the coherent matched filtering within each segment, while during the subsequent incoherent stack–slide combination, the grid is refined in the frequency-derivative dimensions by fixed factors: $\gamma_1$ for $\dot f$ and $\gamma_2$ for $\ddot f$. 

We use two different search set-ups in the low-frequency range (20–500 Hz) and in the high-frequency range (500–1500 Hz). 
For each target and frequency range, we use uniform spacings in $f$, $\dot f$ and $\ddot f$. The resulting setup is characterized by an average template mismatch $\Bar{m}$, which quantifies the expected fractional loss in (squared) signal‑to‑noise due to the finite grid spacing between the true signal parameters and the nearest template. $\Bar{m}$ is estimated via test-signal search and recovery Monte Carlos. All Stage-0 parameters are given in the top block of Table~\ref{tab:search_params_FU}. These search set-ups were determined with the optimisation procedure of \cite{Ming2016}, that maximizes the overall search sensitivity given the available computational budget. 

The search ranges in the first and second order frequency derivative are determined as follows: we begin with the standard assumption that $\dot f  \propto f^{n}$, with $2\leq n\leq 7$ being the braking index governing the frequency evolution. Under this assumption 
\begin{equation}
    \begin{cases}
         \dot{f} = -f/ [(n-1)\tau ] \\
         \ddot{f} = n\dot{f}^2/f.
    \end{cases}
    \label{eqn:perfectPowerLaw}
\end{equation}
We relax this model in two ways: i) we define the ranges of variability of $\dot{f}$ based on the range on $n$, and include in the possible $n$ range non-standard values, i.e. $n\rightarrow \infty$. ii) we further expand the searched waveforms to include ones for which the values of $\dot{f}$ and $\ddot{f}$ searched for a given $f$ are not consistent, in the sense that they correspond to different values of $n$. In particular for the upper bound on $\ddot{f}$ we take $|\dot{f}|=|\dot{f}|_{\textrm{max}}=f/\tau$, which from the first equation in Eqs.~(\ref{eqn:perfectPowerLaw}) corresponds to $n=2$, and then further maximize setting $n=7$ in the second equation of Eqs.~(\ref{eqn:perfectPowerLaw}). This yields
\begin{equation}
    \begin{cases}
        20 \ \text{Hz} \leq  f \leq 1500 \ \text{Hz} \\
         -f/ \tau  \leq  \dot{f} \leq 0 \ \text{Hz/s} \\
         0\,\mathrm{Hz/s}^2 \leq  \ddot{f} \leq 7 {f/\tau^2}. 
    \end{cases}
    \label{eqn:param_space}
\end{equation}
The  $\ddot f = n\,\dot f^2/f$, which is consistent with a power-law behavior,  is within our search range  $\ddot f \le 7f/\tau^{2}$ when $n \in (0, 0.69)\ \cup\ (1.46,\infty)$, and so it lies in our search range for all standard values $n\geq 2$. 
 Our parameter space extension is straightforward to implement and is
motivated by observations that the spin of young pulsars often
deviates from a simple $\dot f \propto f^{n}$ law due to a variety of reasons including timing noise,
glitches, and torque variability (see, e.g., \cite{Vargas:2024itq} and
references therein).

All Stage-0 searches are executed on the Einstein@Home volunteers' platforms, which are implemented on top of the BOINC (Berkeley Open Infrastructure for Network Computing) architecture \citep{Anderson2004,Anderson2006}.  The Einstein@Home project harnesses idle CPU/GPU cycles from ``citizen scientists'' to look for long-lived, weak signals from rotating neutron stars, including continuous gravitational waves. 
Across all three targets we explore $\sim 2\times10^{18}$ waveform templates and use Einstein@Home for about 4 months. The total workload is partitioned into ``work units'' which are tuned to occupy a volunteer host for $\sim 8$ CPU hours or $\sim 20$ GPU minutes each, yielding in aggregate $\sim 8\times10^{6}$ work units.

\begin{figure}[h!tbp]
  \includegraphics[width=\columnwidth]{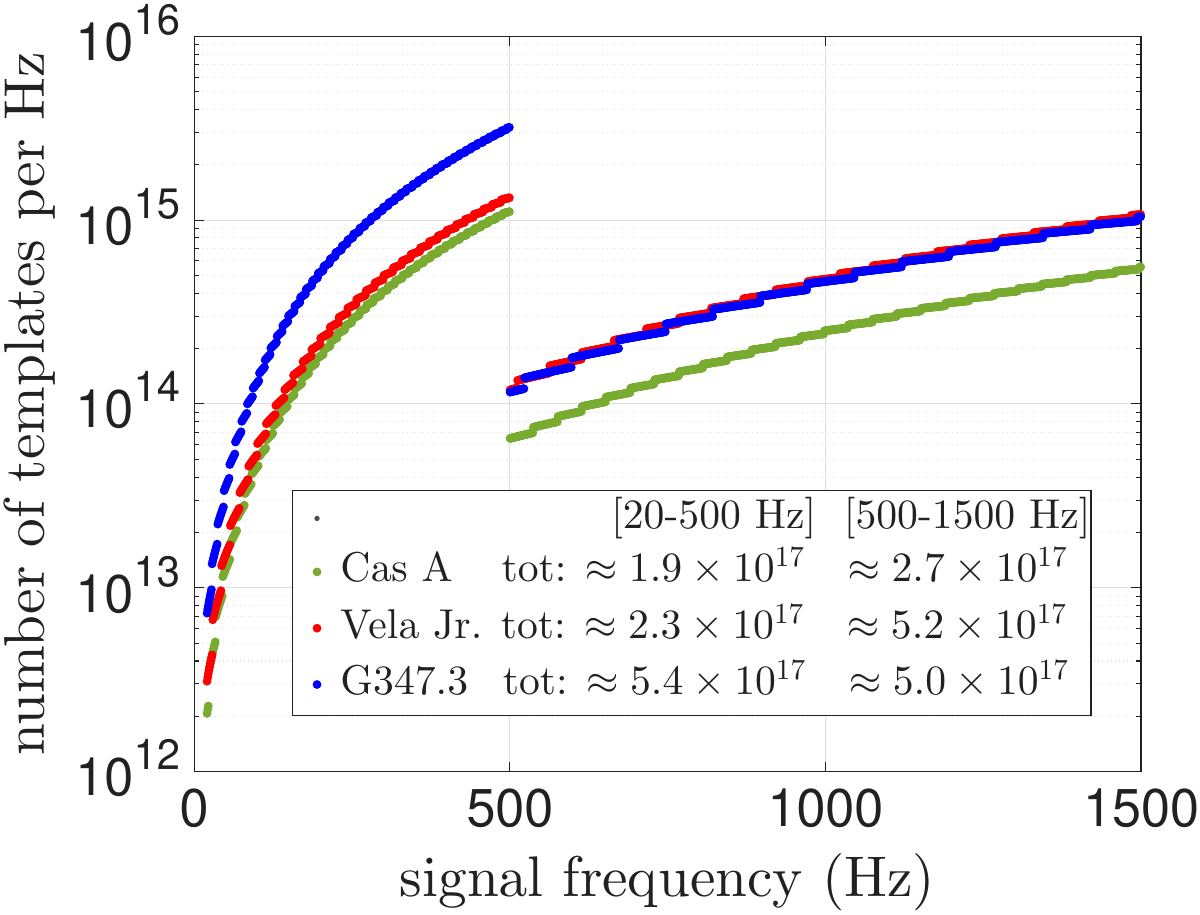}
\caption{Number of templates searched in 1-Hz bands as a function of signal frequency. In the legend we also show the total number of templates searched for each target in the low and high frequency bands.}
\label{fig:HowManyTemplates}
\end{figure}

As the search ranges in $\dot f$ and $\ddot f$ expand with increasing $f$ (Eqs.~(\ref{eqn:param_space})), the number of templates required in a given frequency band grows accordingly. The distribution of templates per Hz band for all targets is displayed in Figure~\ref{fig:HowManyTemplates}. Although the parameter space of the high-frequency search is more than twice larger than that of the low-frequency search,  a very similar number of templates are used in the low and high frequency bands. This is due to the fact that the search set-ups used in the low-frequency band are generally more sensitive than those in the high-frequency band. Similarly, the parameter space searched for Cas A is the largest, due to Cas A's younger age, while the number of templates searched for Cas A is smaller compared to the other two targets, reflecting a less sensitive search. These are all consequences of the optimization scheme described in \cite{Ming2016}, which, for a fixed computing budget, maximizes the overall detection probability by assigning the most sensitive set-ups to those frequency bands and targets that are most promising for a continuous wave detection.

\addtolength{\tabcolsep}{1pt} 
\begin{table*}[t]
    \begin{tabular}{|l c c c c c c c c|}
        \hline\hline
        Stage & $T_{\text{coh}}$ [hr] & $N_{\text{seg}}$ & $\delta f$ [Hz] & $\delta \dot{f}$ [Hz/s] & $\delta \ddot{f}$ [Hz/s$^2$] &$\gamma_1$ & $\gamma_2$& $\Bar{m}$ \\ 
        \hline\hline
         
        {0 - }$\mathrm{E@H}^\mathrm{Cas A}_ \mathrm{20-500 Hz}$& $360$ & $12$ & $4.7 \times 10^{-7}$ & $1.8 \times 10^{-12}$ & $7.3 \times 10^{-19}$  &21 &21 & 17.3\% \\
        \hline
         {0 - }$\mathrm{E@H}^\mathrm{Cas A}_ \mathrm{500-1500 Hz}$& $240$ & $18$ & $7.0 \times 10^{-7}$ & $4.0 \times 10^{-12}$ & $2.5 \times 10^{-18}$ & 13&21 & 32.9\% \\
         \hline
        {0 - }$\mathrm{E@H}^\mathrm{Vela Jr.}_ \mathrm{20-500 Hz}$& $720$ & $6$ & $1.9 \times 10^{-7}$ & $4.5 \times 10^{-13}$ & $2.1 \times 10^{-19}$ &13 &11 & 21.6\% \\
        \hline
         {0 - }$\mathrm{E@H}^\mathrm{Vela Jr.}_ \mathrm{500-1500 Hz}$& $360$ & $12$ & $4.7 \times 10^{-7}$ & $1.8 \times 10^{-12}$ & $7.3 \times 10^{-19}$ &21 & 21& 17.3\% \\
         \hline
        {0 - }$\mathrm{E@H}^\mathrm{G347.3}_ \mathrm{20-500 Hz}$& $1440$ & $3$ & $6.7 \times 10^{-8}$ & $8.7 \times 10^{-14}$ & $2.6 \times 10^{-20}$ &7 &5 & 5.4\% \\
        \hline
         {0 - }$\mathrm{E@H}^\mathrm{G347.3}_ \mathrm{500-1500 Hz}$& $720$ & $6$ & $1.9 \times 10^{-7}$ & $4.5 \times 10^{-13}$ & $2.1 \times 10^{-19}$ &13 & 11& 21.6\% \\
       
        \hline
        \hline
        1 - O3a, {semi-coherent} & $1440$ & $3$ & $6.7 \times 10^{-8}$ & $8.7 \times 10^{-14}$ & $2.6 \times 10^{-20}$ &7 &5 & 5.4\% \\
         \hline
        2 - O3a, {fully coherent} & $4392$  & $1$ & $1.2 \times 10^{-8}$  & $3.8 \times 10^{-15}$  & $3.4\times 10^{-21}$  &   1 & 1& $0.9\%$ \\
         \hline
        3 - O3a+b, {fully coherent}  & $8667$  & $1$ & $6.1 \times 10^{-9}$ & $9.8 \times 10^{-16}$  & $4.4 \times 10^{-22}$ & 1 & 1&  $0.6\%$ \\
          \hline
        4 - O4a, {fully coherent}   & $5688$  & $1$ & $9.3\times 10^{-9}$ & $3.2 \times 10^{-15}$  & $1.3 \times 10^{-21}$ & 1 & 1&  $1.8\%$ \\
        
        \hline
        \hline
    \end{tabular}
    \caption{\label{tab:search_params_FU} Search set-ups for all Stages. The Einstein@Home searches ( ``E@H" ) are different for the three targets and for different frequency ranges whereas the follow-up stages 1-4 use the same set-up for all targets and both frequency ranges. $T_{\text{coh}}$ is the coherent baseline time, $N_{\text{seg}}$ is the number of coherent segments, $\delta f$, $\delta \dot{f}$, and $\delta \ddot{f}$ are the coarse grid-spacings of the templates, $\gamma_1$ and $\gamma_2$ are the refinement factors and $\Bar{m}$ is the average mismatch. }
\end{table*}
\addtolength{\tabcolsep}{1.5pt}

\addtolength{\tabcolsep}{1pt} 
\begin{table*}[t]
    \begin{tabular}{|| l  c c c c c c  ||}
\hline
\hline
Stage & $\Delta f$ [Hz] & $\Delta \dot{f}$ [Hz/s] & $\Delta \ddot{f}$ [Hz/s$^2$] &  $R^a_{\text{thr}}$ & $N_{\text{in}}$ & $N_{\text{out}}$ \\ 
\hline
\hline
\multicolumn{7}{||c||}{\bf{Cas A 20-500 Hz}} \\
\hline
E@H &  full range  & full range & full range  & - & $1.9\times10^{17}$  &$4.8\times10^{6}$\\
1 & $3.2\times 10^{-7}$  & $7.1\times 10^{-14}$  & $3.5\times 10^{-20}$ & 2.8 & $4.8\times10^{6}$  & $385,130$  \\
2 & $6.9\times 10^{-8}$  & $1.4\times 10^{-14}$ & $8.8\times 10^{-21}$  & $8.7$ & $385,130$  & $6,881$  \\
3 & $2.8\times 10^{-8}$ & $6.3\times 10^{-15}$ & $5.0\times 10^{-21}$  & $15.5$ & $6,881$ & $0$  \\
\hline
\multicolumn{7}{||c||}{\bf{Cas A 500-1500 Hz}} \\
\hline
E@H &  full range  & full range & full range  & - & $2.7\times10^{17}$  & $1.0\times10^{7}$ \\
1 & $7.7\times 10^{-7}$  & $4.3\times 10^{-13}$  & $1.2\times 10^{-19}$ & 3.7 & $1.0\times10^{7}$   & $586,883$  \\
2 & $7.0\times 10^{-8}$  & $1.2\times 10^{-14}$ & $7.0\times 10^{-21}$  & $10.8$ & $586,883$  & $24,000$  \\
3 & $2.6\times 10^{-8}$ & $7.6\times 10^{-15}$ & $4.3\times 10^{-21}$  & $21.4$ & $24,000$ & $0$  \\
\hline
\multicolumn{7}{||c||}{\bf{Vela Jr. 20-500 Hz}} \\
\hline
E@H &  full range  & full range & full range  & - & $2.3\times10^{17}$  &$4.8\times10^{6}$ \\
1 & $1.9\times 10^{-7}$  & $3.8\times 10^{-14}$  & $1.6\times 10^{-20}$ & 1.6 & $4.8\times10^{6}$  & $1,668,988$  \\
2 & $7.6\times 10^{-8}$  & $2.2\times 10^{-14}$ & $9.0\times 10^{-21}$  & $4.9$ & $1,668,988$  & $56,920$  \\
3 & $2.9\times 10^{-8}$ & $6.2\times 10^{-15}$ & $5.2\times 10^{-21}$  & $7.9$ & $56,920$ & $1$  \\
4 & $2.9\times 10^{-5}$ & $4.3\times 10^{-13}$ & $3.2\times 10^{-21}$  & $4.7$ & $1$ & $0$  \\
\hline
\multicolumn{7}{||c||}{\bf{Vela Jr. 500-1500 Hz}} \\
\hline
E@H &  full range  & full range & full range  & - & $5.2\times10^{17}$  &$1.0\times10^{7}$ \\
1 & $4.1\times 10^{-7}$  & $7.7\times 10^{-14}$  & $3.7\times 10^{-20}$ & 2.8 & $1.0\times10^{7}$   & $759,939$  \\
2 & $7.3\times 10^{-8}$  & $1.1\times 10^{-14}$ & $7.3\times 10^{-21}$  & $8.8$ & $759,939$  & $9,959$  \\
3 & $3.6\times 10^{-8}$ & $6.7\times 10^{-15}$ & $5.9\times 10^{-21}$  & $14.6$ & $9,959$ & $0$  \\
\hline
\multicolumn{7}{||c||}{\bf{G347.3 20-500 Hz}} \\
\hline
E@H &  full range  & full range & full range  & - & $5.4\times10^{17}$  & $4.8\times10^{6}$ \\
1 & -  & -  & - & - & -   & -  \\
2 & $8.2\times 10^{-8}$  & $2.5\times 10^{-14}$ & $1.0\times 10^{-20}$  & $2.5$ &$4.8\times10^{6}$  & $488,766$  \\
3 & $3.9\times 10^{-8}$ & $8.9\times 10^{-15}$ & $6.1\times 10^{-21}$  & $4.0$ & $488,766$ & $13$  \\
4 & $3.3\times 10^{-5}$ & $5.0\times 10^{-13}$ & $3.7\times 10^{-21}$  & $2.0$ & $13$ & $1$  \\
\hline
\multicolumn{7}{||c||}{\bf{G347.3 500-1500 Hz}} \\
\hline
E@H &  full range  & full range & full range  & - & $5.0\times10^{17}$  & $1.0\times10^{7}$ \\
1 & $1.9\times 10^{-7}$  & $4.5\times 10^{-14}$  & $1.8\times 10^{-20}$ & 1.5 & $1.0\times10^{7}$   & $3,912,806$  \\
2 & $8.2\times 10^{-8}$  & $2.3\times 10^{-14}$ & $1.1\times 10^{-20}$  & $4.8$ & $3,912,806$  & $171,539$  \\
3 & $3.6\times 10^{-8}$ & $8.4\times 10^{-15}$ & $6.5\times 10^{-21}$  & $8.0$ & $171,539$ & $0$  \\
\hline
\hline
\end{tabular}
 \caption{\label{tab:results_FU} The quantities $\Delta f$, $\Delta\dot f$, and $\Delta\ddot f$ denote, for each candidate, the one-sided search range used at Stages~1-4. 
 For the initial Einstein@Home (E@H) search survey, the search ranges are full frequency range and its first two derivatives ranges as described in Eqs.~(\ref{eqn:param_space}).
We use $N_{\mathrm{in}}$ for the number of candidates actually searched at Stages~1-4 (and for the total template count in the Einstein@Home run), and $N_{\mathrm{out}}$ for the number of survivors promoted from each stage. The quantity $R^{a}_{\mathrm{thr}}$ is the cut defined in Eq.~(\ref{eqn:R_athr}) (applied for $a=1,2,3,4$). Entries from the Einstein@Home survey are chosen via the pixeling scheme, so no $R_{\mathrm{thr}}$ is applied there. For G347.3 in the 20-500 Hz band, the Stage-1 set-up was used in Stage-0, so the first follow-up used directly the Stage-2 set-up. For all three targets in the 500-1500 Hz band,  Stage-4 was not executed, because we vetoed all candidates at Stage-3 and no more candidates need to be verified at Stage-4.}
\end{table*}
\addtolength{\tabcolsep}{1.5pt}


\section{\label{sec:FU} Hierarchical follow up searches}

There are in total four stages of searches.  Stage-0 is the wide semi-coherent Einstein@Home search described above, while the  subsequent Stages~1-4 are hierarchical follow-up searches, in which we re-search localized neighborhoods around the most significant outliers selected from the previous stages. All the follow-up searches are performed on the in-house Atlas super-computing cluster \citep{Atlas}.

\subsection{Stage-0 and candidate selection}

After a work-unit is complete, the host returns the top-ranking $10^{5}$ results to the Einstein@Home project. Their ranking on the volunteer hosts is based on the line-robust statistic $\hat{\beta}_{\mathrm{S/GLtL}}$, whose larger values indicate that the data are more consistent with signal plus noise than with noise alone. Over all searches these add up to $\sim 8\times 10^{11}$ results. 

A pixeling procedure further selects candidates to follow up from these $\sim 8\times 10^{11}$ results. The pixeling procedure was first used in the context of Einstein@Home in the O2 searches \citep{ming2025}. Pixeling consists in selecting a fixed number of top candidates from equal-extent sub-volumes of the searched parameter space. This yields a more even selection of candidates across the parameter space compared to threshold-based or top-ranking methods or their sophisticated combinations and variations. In the context of Einstein@Home directed searches the pixeling procedure has several advantages compared to the traditional clustering methods \citep{Singh:2017kss,Beheshtipour:2020zhb,Beheshtipour:2020nko,den_cluster}. Simulation results show that in undisturbed bands it results in $<1\%$ loss in the final search sensitivity compared with highly optimised clustering methods, e.g. \cite{den_cluster}, while in disturbed frequency bands, it results in much better search sensitivity (30\% better in sensitivity in \cite{ming2025}). Very importantly, the pixeling procedure is easy to implement and does not require as much parameter tuning as the clustering methods, which is also computationally costly. 

In this search, we apply the pixeling procedure as follows:
(i) For each target, we partition the full searched frequency range into $N_{\textrm{bands}}$ bands with equal width 0.5 Hz. 
(ii) Within each 0.5-Hz band we tile the $(f,\dot f)$ plane into $N_p=100 \times 10$ pixels, where 100 pixels are along the 0.5-Hz frequency band and 10 pixels are along the spin-down $f/\tau$ direction.
(iii) We retain the 5 top-ranking Stage-0 results from each pixel, except for an extremely small cohort having too low detection statistic values, as we will explain in the next section. We call these selected results, {\it{candidates}}.
 
The choice of  {\it{5}} top-ranking results is dictated by the available computational budget for the follow-up. These choices result in  $\sim 45$ million candidates to follow up.

\subsection{Coherence growth and target signal population}
The principle of our follow-ups is to lengthen the coherent integration so that signals accumulate more significance than noise fluctuations or coherent disturbances. We introduce the Stage-$a$ growth ratio 
\begin{equation}
     R^a \equiv \frac{2 \Bar{\mathcal{F}}_r^{\text{ Stage-a}} - 4}{2 \Bar{\mathcal{F}}_r^{\text{ Stage-0}} - 4} ~~~{\textrm{for a }}=1,2, 3, 4
     \label{eqn:R_a}
 \end{equation}
to measure the increase with respect to Stage-0. Because $2 \Bar{\mathcal{F}}_r^{\text{ Stage-a}} - 4$ is the expected value of the non-centrality parameter, for a signal it approximately scales with the coherent duration $T_{\text{coh}}$, so $R^a$ is expected to grow proportionally with $T_{\text{coh}}^a$ \citep{JKS1998}.

At every stage we impose a cutoff $R^{a}_{\mathrm{thr}}$ and reject candidates that fail to grow sufficiently:
\begin{equation}
R^{a}<R^{a}_{\mathrm{thr}} \Longrightarrow \text{candidate rejected.}
\label{eqn:R_athr}
\end{equation}
These thresholds are tuned based on the results of searches performed on data containing fake signals drawn from our target population: thousands of test signals are added to the data, the full pipeline is run, and the resulting $R^{a}$ values are recorded. We choose $R^{a}_{\mathrm{thr}}$ at each stage and for each target so that fewer than $0.01\%$ of test signals are falsely discarded, ensuring a conservative selection while efficiently pruning noise outliers.

The parameters of the test signal population are chosen to be representative of the real astrophysical signal population that we want to detect.
For each supernova remnant, the signals are placed at the known sky position of the source, while their frequency and spin-downs parameters are drawn randomly from uniform distributions over the corresponding search priors in frequency and its derivatives.

The orientation angles are sampled uniformly, with $\cos\iota:[-1,1]$ and $\psi:[-\pi/4,\pi/4]$ (the latter reflecting the $\pi/2$ periodicity of the polarization angle’s fundamental domain, see Eqs.~(10) and (11) of  \cite{JKS1998}). Signal strengths are restricted to narrow intervals centered on the target $h_0$ values which are expected to yield approximately $90\%$ detection efficiency. 

Less than 0.005\% of the 45 million candidates have $2 \Bar{\mathcal{F}}_r^{\text{ Stage-0}} $ values lower than those of our test signal population. 
These candidates come from disturbed bands in the low-frequency region and they have extremely low  $2 \Bar{\mathcal{F}}_r^{\text{ Stage-0}}$ values because of:
(i) the trials factor due to frequency range: the number of templates searched per Hz in the low frequency regions is much smaller than at higher frequencies (see Figure~\ref{fig:HowManyTemplates} and Eqs.~(\ref{eqn:param_space})), so even weak candidates can reach the Einstein@Home top list.
(ii) the trials factor due to top-list saturation: because the Einstein@Home results are pure top-lists (no pixeling), when a band is affected by a disturbance, most candidates come from the disturbed part of the 50-mHz band, leaving pixels outside the disturbed regions with very few candidates. Our candidates come from undisturbed pixels within a disturbed band, which means they are the top 5, picked among a very small set.

As explained above, the follow-up procedure was tuned on test-signal populations. This means that the follow-up is not reliably characterized for signals weaker than those of the test-signal populations. We hence
exclude from the follow-up very weak candidates by imposing a lower cutoff threshold
\begin{equation}
2\Bar{\mathcal{F}}_r^{\text{ Stage-0}}< 2 \Bar{\mathcal{F}}_\mathrm{thr}^{\text{ Stage-0}} \Longrightarrow \text{candidate rejected,}
\label{eqn:2Fstage0-thr}
\end{equation}
with  $2 \Bar{\mathcal{F}}_\mathrm{thr}^{\text{ Stage-0}}=$
9.7, 13.3, and 20.2, for Cas A, Vela Jr., and G347.3 respectively. These threshold values are the lowest recorded from our studies on the test-signal population for each target.

The follow-up search set-ups are the same for all targets.
From Stage-1 through Stage-3, we only use O3 data and we progressively lengthen $T_\mathrm{coh}$ while tightening the average mismatch $\Bar m$, which increases the expected excess square of signal-to-noise ratio.
At Stage-4, we follow up the remaining candidates with the newly released O4a data.
The grid spacings and corresponding $\bar{m}$ of the follow-up stages are summarized in the lower block of Table~\ref{tab:search_params_FU}.

The search-and-recovery simulations on the target population also determine how wide a neighborhood parameter space must be searched around each candidate at every stage. This neighborhood space—commonly referred to as the \emph{containment region}—is specified, for Stage-$a$, by the distances between recovered candidate parameters and real signal parameters, within which $> 99\%$ of the target-population candidates lie.
The containment regions $\Delta f^a$, $\Delta \dot{f}^a$,  $\Delta \ddot{f}^a$ are the extents to the right and left of the template parameters of a Stage-$(a-1)$ candidate, searched at Stage-$a$.  From Stage-1 to Stage-3, the next stage follow-up always uses a longer $T_\text{coh}$ and finer grid spacings, which lead to a smaller containment region. 
At Stage-4 the search region is the containment region from the full O3 search evolved to the time of the O4a data (see the details of Eq.~(10) in \cite{ming2024a}).
The search ranges adopted at each stage are summarized in Table~\ref{tab:results_FU}.

\begin{figure}[]
\centering
\subfigure{
    \includegraphics[width=1.0\columnwidth] {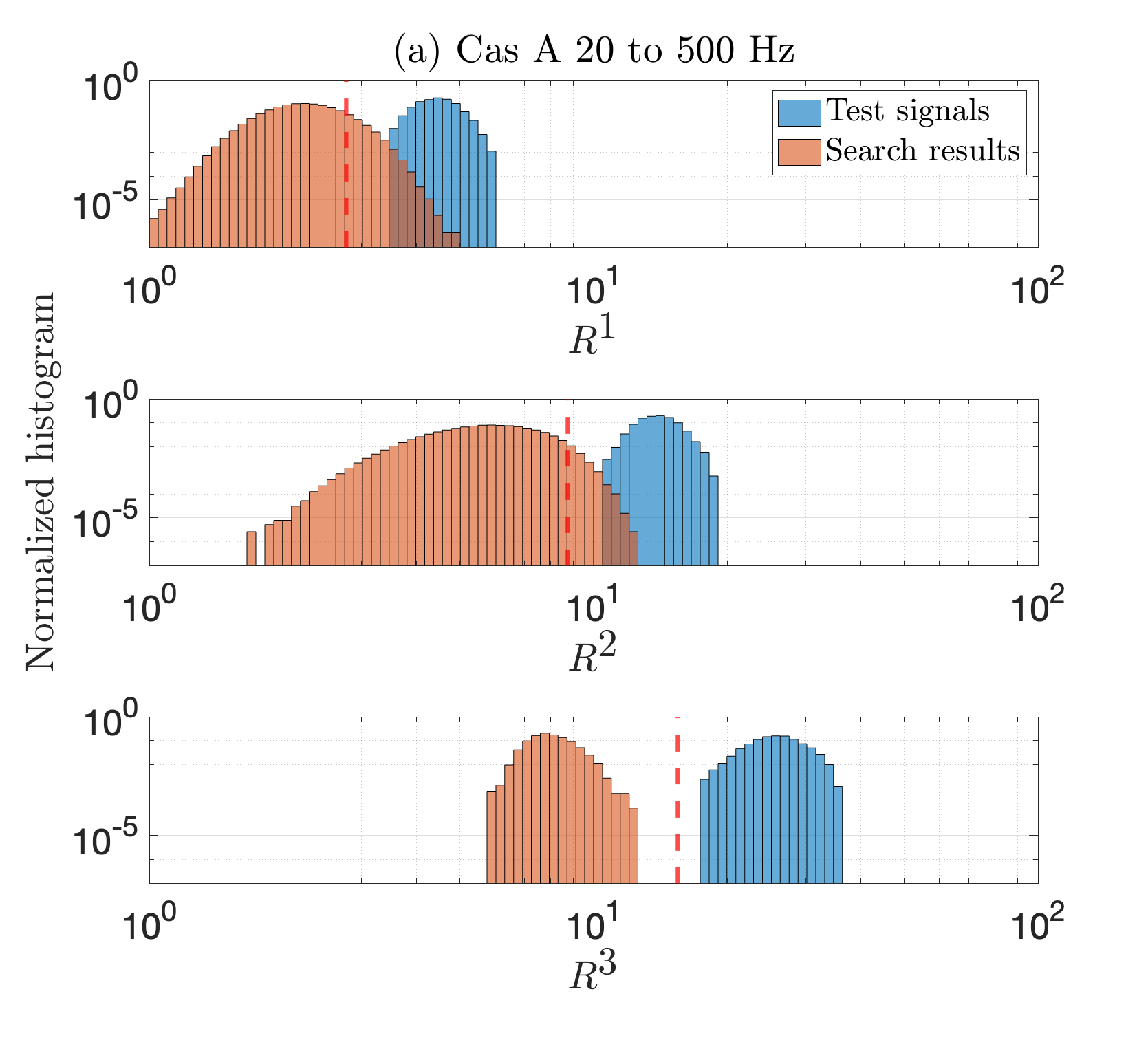}
    \label{fig:casaR1}}
    \subfigure{
    \raisebox{0mm}{
       \includegraphics[width=1.0\columnwidth] {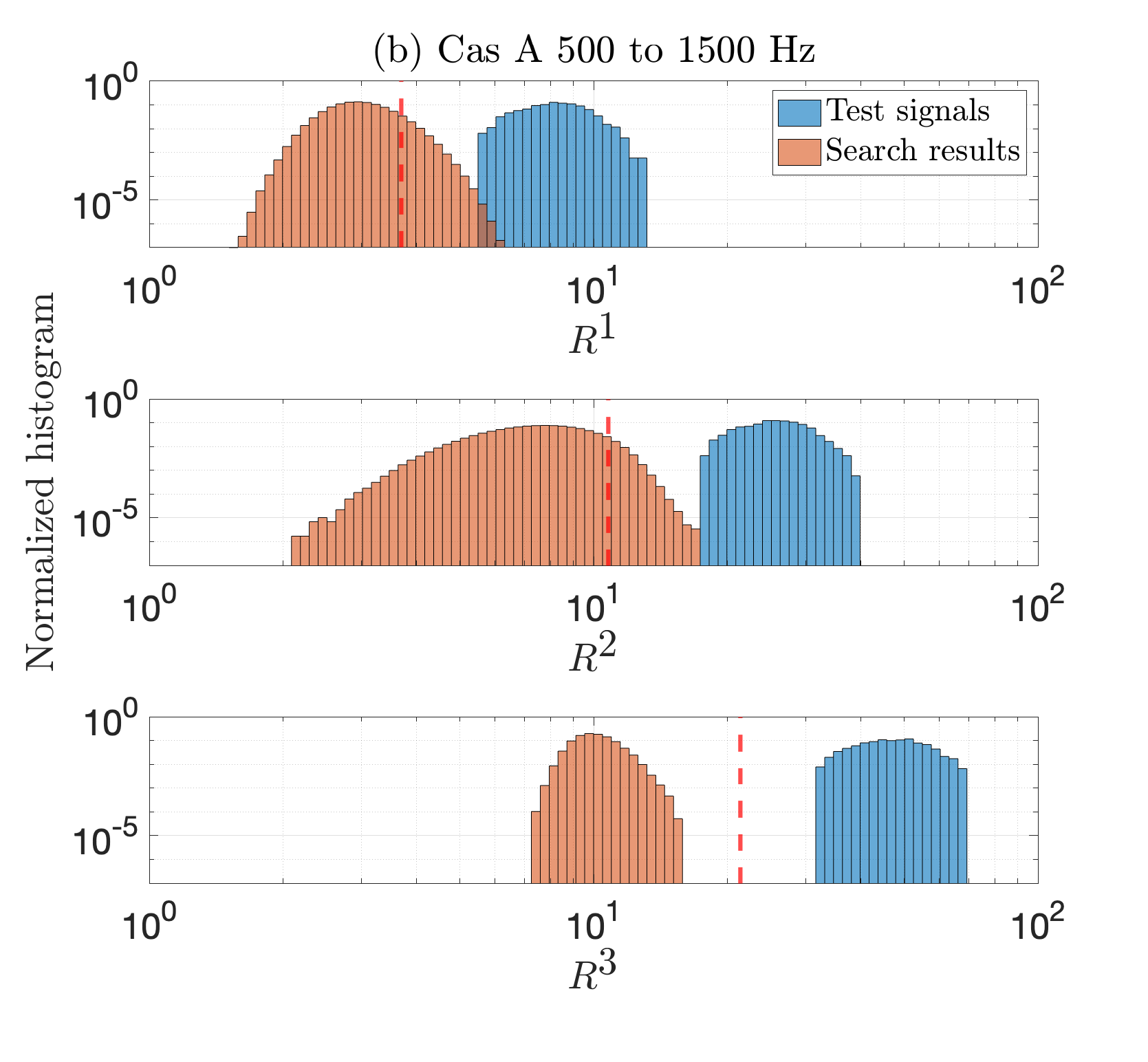}
          \label{fig:casaR2}}}
    \caption{Distributions of $R^{a}$ from all the Cas A searches. In orange color are the distributions of the growth ratio $R^{a}$ for the search candidates; in blue are the results for the test-signals drawn from the target population. The vertical dashed line indicates the per-stage threshold $R^{a}_{\mathrm{thr}}$. As the hierarchical follow-ups progress, noise-dominated outliers increasingly cluster at lower $R^{a}$, while candidates associated with signals shift toward larger values, yielding a progressively clearer separation between the two populations.}
    \label{fig:casaR} 
\end{figure}

\begin{figure}[]
\centering


\subfigure{
    \includegraphics[width=1\columnwidth] {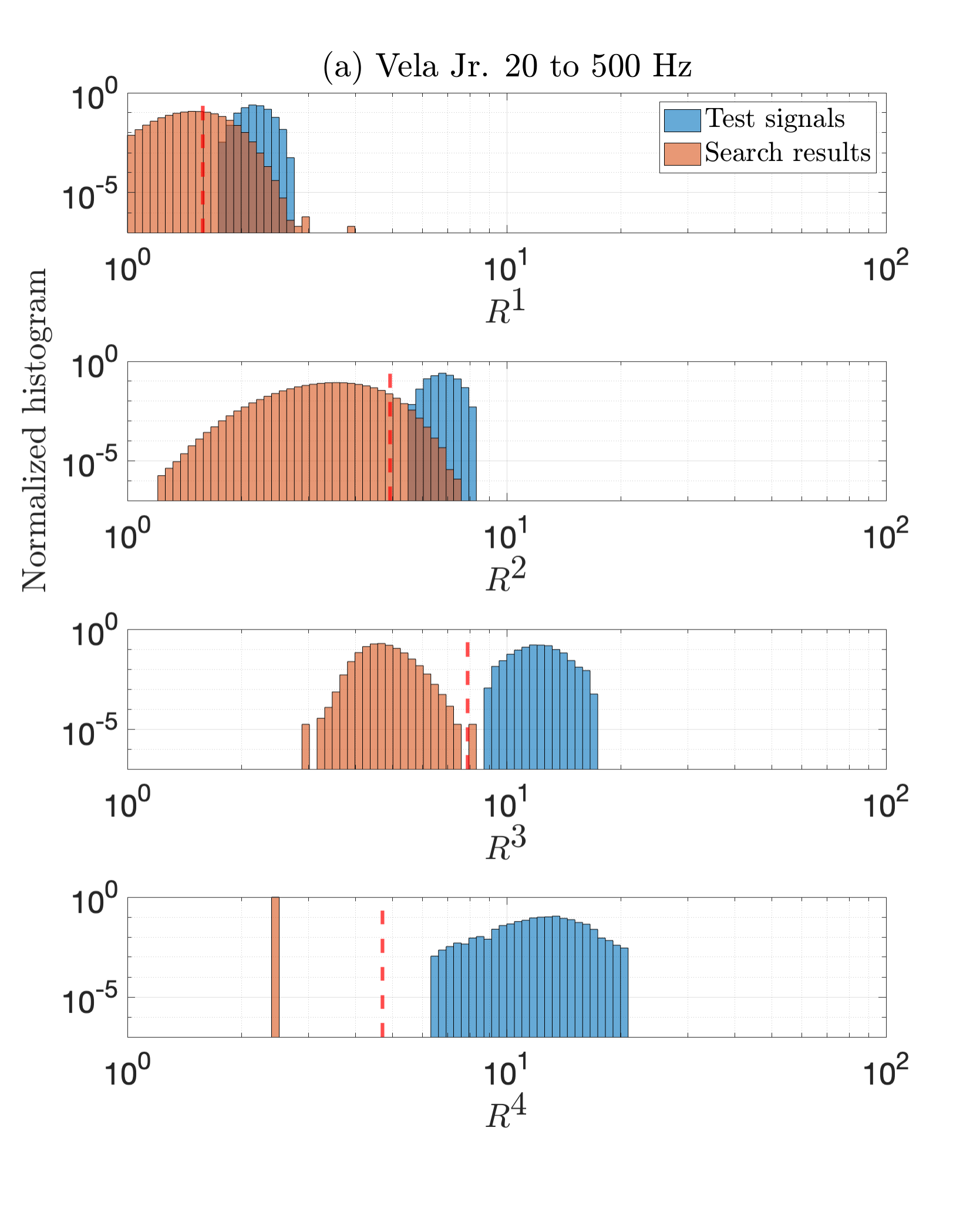}
    \label{fig:velaR1}}
    \subfigure{
       \includegraphics[width=1\columnwidth] {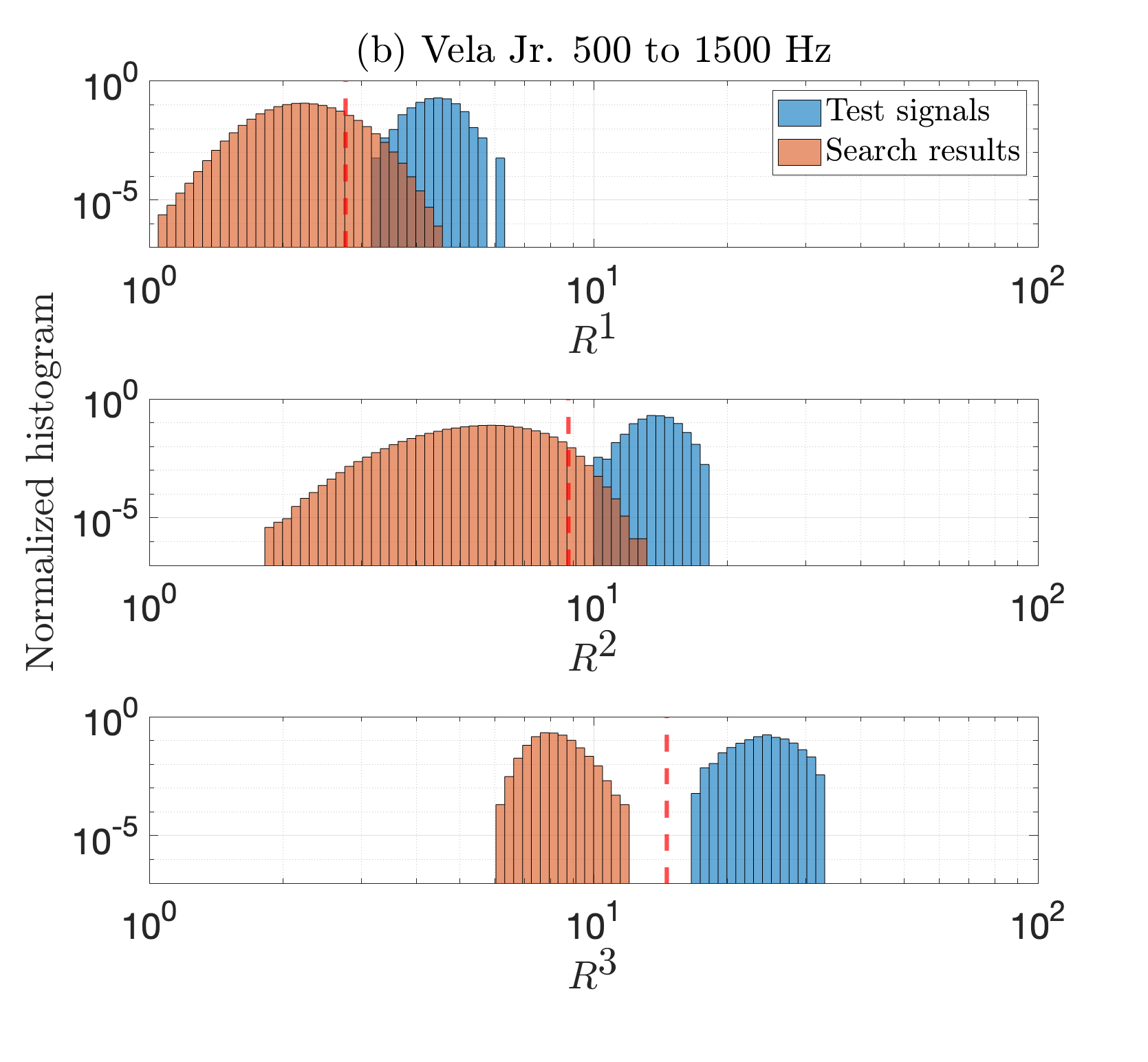}
          \label{fig:velaR2}}
          
       \vspace*{-8pt} 
        
    \caption{Distributions of $R^{a}$ from all the Vela Jr. searches.}
    \label{fig:velaR} 
\end{figure}


\begin{figure}[]
\centering
\vspace*{-1pt}
\subfigure{
    \raisebox{-19.65mm}{
      \includegraphics[width=1\columnwidth] {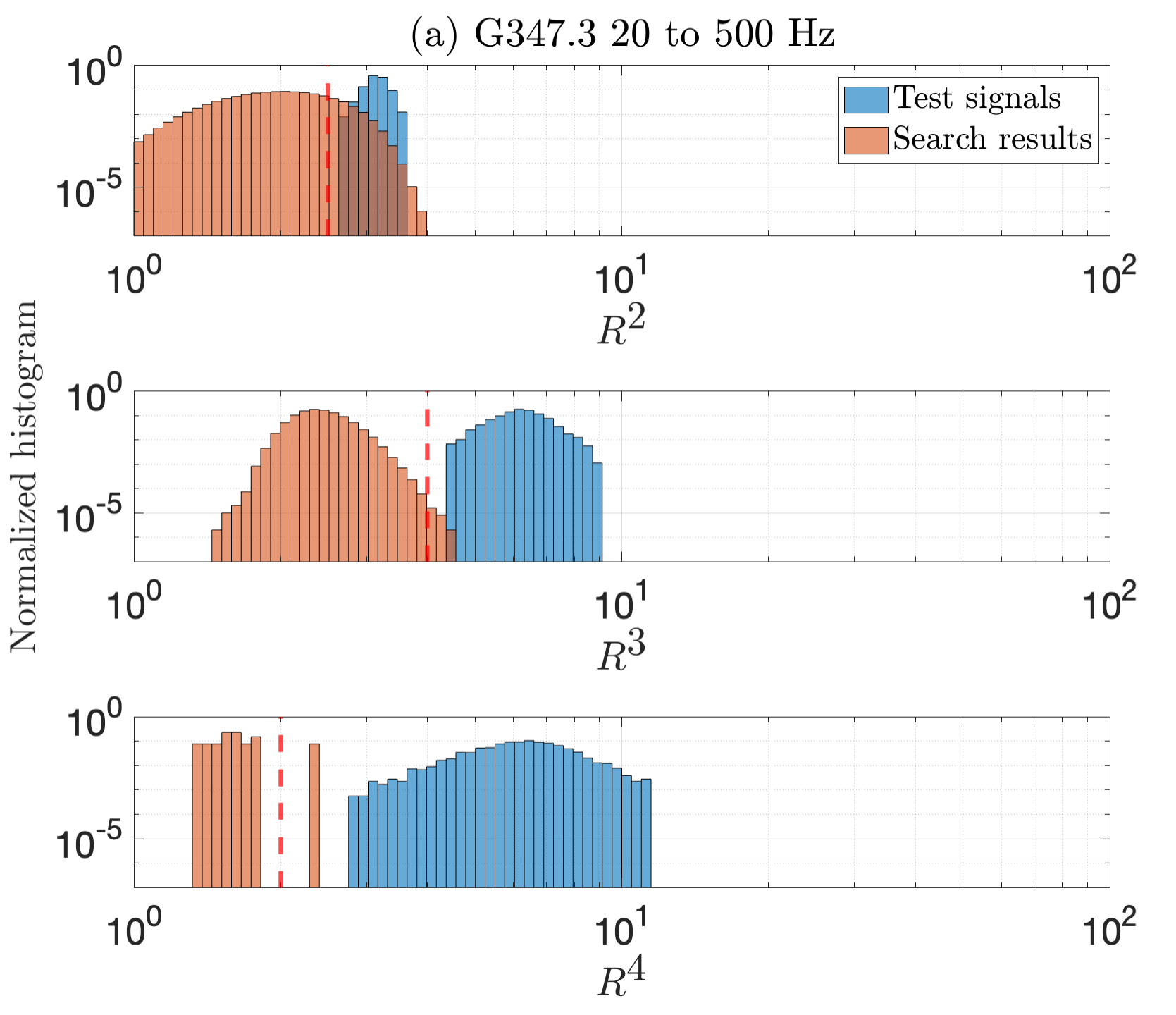}
    \label{fig:3473R1}}}
    \subfigure{
       \includegraphics[width=1\columnwidth] {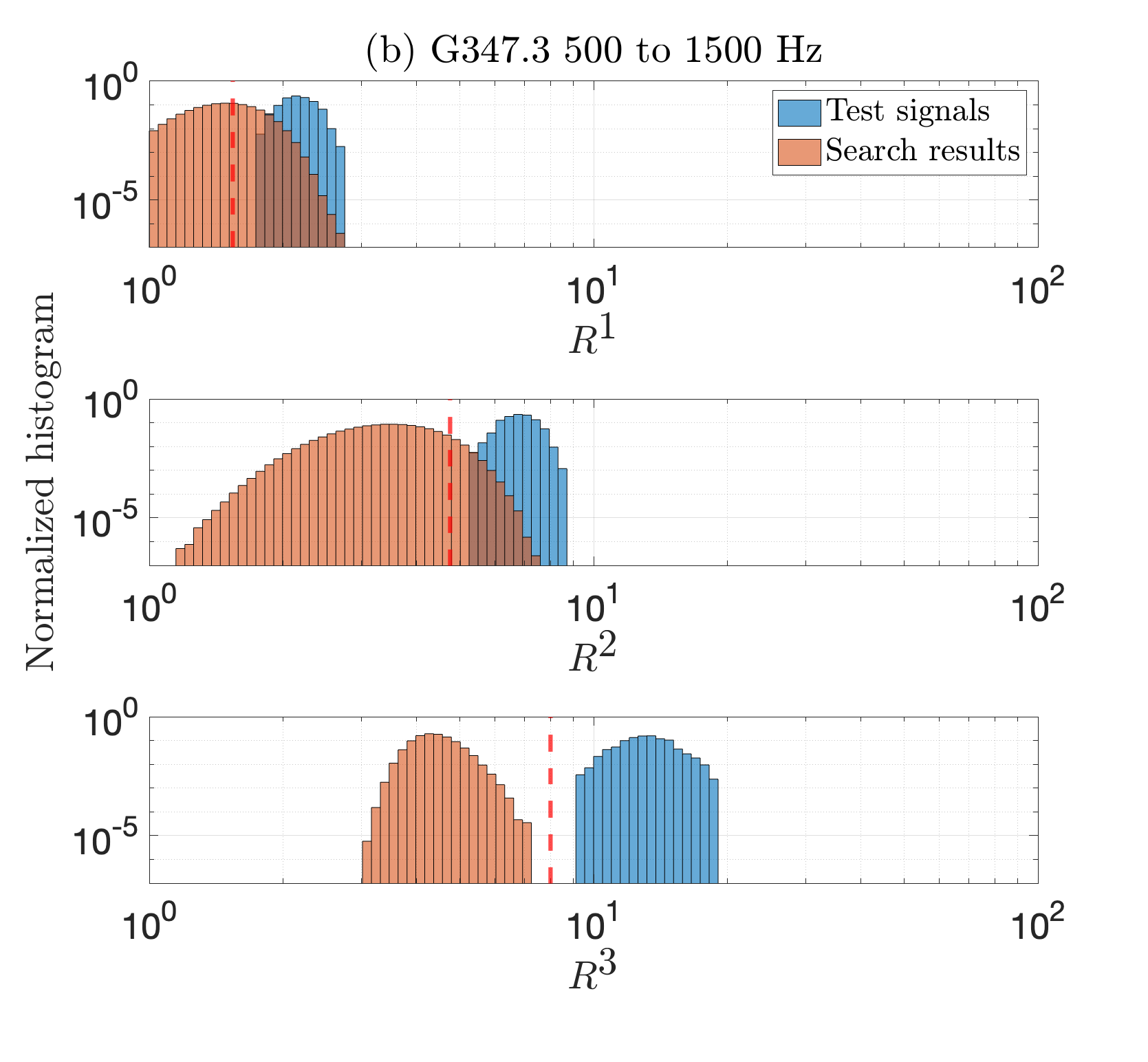}
          \label{fig:3473R2}}
          
          \vspace*{-8pt} 
         
    \caption{Distributions of $R^{a}$ from all the G347.3 searches. }
    \label{fig:3473R} 
\end{figure}

\subsection{\label{subsec:FU_stage_1} Stage-1 follow-ups}


Stage-1 follows up the candidates selected with pixeling. We employ the search set-up used at Stage-0 for the low-frequency G347.3 search, with $T_\text{coh}=1440$ hours (close to two months), and grid spacings given in the seventh row of Table~\ref{tab:search_params_FU}. This means that about 5 million pixeling candidates from the  G347.3 low-frequency Stage-0 search go straight to the Stage-2 follow-up. So in total, with Stage-1 we follow up $\sim 40$ million candidates.

In general, finer grids and longer $T_\text{coh}$ in Stage-0, yield smaller uncertainties in candidate parameters, i.e. smaller search ranges for Stage-1. For example, in Table~\ref{tab:results_FU}, the Stage-1 $\Delta f^\mathrm{Cas A}_ \mathrm{500-1500 Hz}  \approx 8\times10^{-7}$ Hz is the uncertainty around candidates from a Stage-0 search which has a $T_\text{coh}=240$ hrs and $\Bar{m}\approx 33\%$. 
The Stage-1 $\Delta f^\mathrm{Vela Jr.}_ \mathrm{20-500 Hz}$ stems from a Stage-0 search with a much longer $T_\text{coh}=720$ hrs and a finer grid $\Bar{m}\approx 22\%$, resulting in $\Delta f^\mathrm{Vela Jr.}_ \mathrm{20-500 Hz}\approx 2\times10^{-7}$ Hz, which is about 4 times smaller than that for the Stage-1 Cas A high-frequency search. 

The number of candidates $N_{\mathrm{in}}$ and $N_{\mathrm{out}}$ respectively entering and surviving each follow-up stage is given in Table~\ref{tab:results_FU}.
Even though the number of candidates $N_{\mathrm{in}}$ being fed to the Stage-1 follow-up is the same for all searches of the same frequency range, the number of surviving candidates $N_{\mathrm{out}}$ is vastly different. For example, $N_{\mathrm{in}}$ of Cas A and  Vela Jr. at 20-500 Hz are both 4.8 million, but the number of surviving candidates for Vela Jr. is more than 4 times larger than that of Cas A. 
Since the two Stage-1 searches are identical, the reason for this difference lies in the nature of the input Stage-0 candidates: the Cas A ones stem from a shorter $T_\text{coh}=360$ hours, compared to the $720$ hours of the Stage-0 Vela Jr. search. The shorter coherent baseline of the Cas A search results in a larger difference in coherence time between the Stage-0 and Stage-1 searches, which produces a clearer separation between the $R^1$ of the test-signals and the noise-dominated ones of the search results, shown in the top of Figure~\ref{fig:casaR1} and Figure~\ref{fig:velaR1}. This is also illustrated by the different vetoing threshold $R^1_\mathrm{thr}$ values (also given in Table~\ref{tab:results_FU}):  for Cas A $R^1_\mathrm{thr}=2.8$ whereas for Vela Jr. it is 1.6, while both thresholds correspond to $<0.01\%$ false dismissal rate for signals. 

The Stage-1 low-frequency searches veto about 65\% of the Stage-0 Vela Jr. candidates and about 92\% of the Cas A candidates.
The higher frequency Stage-1 searches veto 94\%, 92\% and 61\% of the candidates from Cas A, Vela Jr. and G347.3, respectively. The detailed distributions of $R^1$ and the corresponding $R^1_\mathrm{thr}$ are shown in Table~\ref{tab:results_FU} and in the top plots of Figures~\ref{fig:casaR},  \ref{fig:velaR} and \ref{fig:3473R2}.

%
\subsection{\label{subsec:FU_stage_2} Stage-2  follow-ups}
%
With Stage-2 we follow up the Stage-1 survivors with a \emph{fully coherent} search spanning the entire O3a data set, which triples the duration of the coherent time baseline of the previous stage. 
Together with much finer grids (the average mismatch is more than 5 times lower than the one of the previous stage), the long coherence baseline yields an increasingly clearer separation between the $R$ values from the noise-dominated real data and the search-and-recovery of the target signal population (see Figures~\ref{fig:casaR} to \ref{fig:3473R}).

Stage-2 overall vetoes over 90\% of the Stage-1 survivors, with an average veto rate of 95\%. However, there are still tens (or hundreds) of thousands of Stage-2 survivors for each target. The $R^2_\mathrm{thr}$ and the exact number of survivors $N_\mathrm{out}$ are shown in Table~\ref{tab:results_FU}.

%
%
\subsection{\label{subsec:FU_stage_3} Stage-3 follow-ups}

The candidates surviving Stage-2 are further investigated with a fully coherent search that uses the combined O3a and O3b data. The time span of this data set is just under a year, further doubling the coherent time with respect to the previous stage.
The distributions of $R^3$ for the search candidates are further separated from those of the signals, as shown in the bottom subplots of Figures~\ref{fig:casaR} to \ref{fig:3473R}.

After Stage-3 all candidates from all the high frequency searches are rejected. The low-frequency Cas A search candidates are also all rejected. Only one candidate remains from the low frequency Vela Jr. search and 13 candidates from the low frequency G347.3 search.

%
%
\subsection{\label{subsec:FU_stage_4} Stage-4 O4a follow-ups}

We verify the surviving candidates from the previous stage with a fully coherent search using the newly released O4a data set, which at the time of writing is the newest data set publicly available. This has a time span of about 237 days, which is $\approx$ 34\% shorter than the data used in the previous stage. The data are, however, $\approx$ 50\% more sensitive than the data used in the previous stage. 

The distributions of $R^4$ for the search candidates and for the target signals are shown in the bottom subplots of Figures~\ref{fig:velaR1} and \ref{fig:3473R1}. 

The Vela Jr. candidate is rejected at this stage, and 12 out of the 13 G347.3 candidates are also rejected as a result of the Stage-4 search.  The surviving candidate is discussed in the next Section.

\subsection{\label{subsec:G347survivor} The surviving candidate}

The only candidate that survives the Stage-4 search comes from the G347.3 low frequency search. Its $R^4=2.3$  is marginal with respect to the test signal population, but it is still above the threshold. 

The slightly negative value of the line-robust detection statistic $\hat{\beta}_{\mathrm{S/GL}}=-0.3$ in Stage-4 
perhaps indicates a slight imbalance among the two detectors in the contribution to the detection statistic, however direct inspection of the single detector contributions does not unveil any significant discrepancy: $2\mathcal{F}_{\textrm{H1}}\simeq 23$
and $2\mathcal{F}_{\textrm{L1}}\simeq 24 $
for the Hanford and Livingston data respectively. 

The O3 and O4a amplitude spectral density data used in the searches are quite pristine at the frequencies of interest for this candidate: no known line was reported in either detector \citep{o3_linefile,o4_linefile} and the amplitude spectral density estimates appear undisturbed, albeit not stationary.  
The Einstein@Home raw results from the deep low-frequency search \citep{brian2025a} do not show any sign of contamination in this frequency band. The all-sky full O3 atlas results \citep{Dergachev:2025ead} show signs of contamination at the sky position of G347.3 from elevated signal-to-noise ratios at the poles, which are the tell-tale of a stationary line. A preliminary investigation using on O4a data the same Falcon pipeline that produces the atlas, reveals at the parameters of our O4a candidate contamination at the position of G347, from elevated signal-to-noise ratio values that peak somewhere else in the sky. On the other hand, these Falcon searches use a different strain channel than what is used here, one where no line-cleaning was carried out by LIGO before the data release.

The Stage-4 search investigates $\sim 10^7$ templates for each candidate. With so many templates noise alone produces the bulk of $2\mathcal{F}$ follow-up results in the range 34-38. Our candidate has a $2\mathcal{F}\simeq 45$, which corresponds to a Gaussian p-value of about 0.75\%. But having carried out 13 follow-ups, the actual false-alarm probability associated with this occurrence is around 10\%. This makes the candidate not highly significant.

The candidate's frequency is $\approx$ 31.7 Hz and it has frequency-derivative values $\dot{f}\approx -3.6\times 10^{-10}$ Hz/s and $\ddot{f}\approx 7.7\times 10^{-20}$ Hz/s$^2$. If the system was spinning down solely due to gravitational-wave emission, at its distance of $1.3$ kpc, $h_0\approx 2.1\times 10^{-24}$ -- this is the spin-down amplitude of our candidate. 
The $h_0$ amplitude estimates \citep{Ashok2024} based on the full O3 data are very consistent with those based on the O4a data, and yield posteriors with significant support at $h_0\simeq 6\times 10^{-26}$ -- 35 times below the spin-down amplitude, corresponding to a fairly large ellipticity of $\approx 7\times 10^{-5}$. Should a signal like this already have been identified? The short answer is ``no": Based on their upper limits ranging between $\approx [7\times 10^{-25}, 2\times 10^{-24}]$ depending on the orientation of the neutron star, \cite{Salvadore2025} would likely not have detected this signal. 
Our previous search on O2 data \citep{ming2024a} at these frequencies has better sensitivity, with upper limits in the $3\times 10^{-25}$ range, but the band where this candidate falls was excluded from the upper limit statements due to the presence of disturbances. The most sensitive all-sky surveys \citep{brian2025b,Dergachev:2025ead} also attain at this frequency stringent $h_0$ upper limits between $[2.7-2.9]\times 10^{-25}$, but would have likely missed a signal this weak.

Closer inspection of phase-parameter consistency between the results on O3 and O4a data is not convincing. We repeat Stage-3 and Stage-4 using the Bayesian follow-up method of \cite{Martins:2025jnq}. We find that the O4a frequency posterior is not much different from its prior, which is the O3 posterior. If a third-order frequency-derivative parameter is introduced, all the O4a posteriors are more informative than their priors (the O3 posterior). Overall, the O4a evidence decreases with respect to the O3 evidence.

Investigations on new data will aid in clarifying the nature of this candidate. Such ``new" data exist and would be optimal for this purpose, because they are close in time to the O4a data set, hence the propagated parameter uncertainties would be limited. We are, however, unable to carry out these further investigations because these data are not publicly accessible at the time of writing.

\section{\label{sec:results} Results}


\subsection{Upper limits on the gravitational wave amplitude}
\label{sec:h0ULs}

We determine frequentist $90\%$-confidence upper limits on the gravitational-wave amplitude $h_0^{90\%}(f)$ in every $0.5$ Hz band, consistent with our search results.  $h_0^{90\%}(f)$ is the strain amplitude for which $90\%$ of signals drawn from our search space and with frequency in a half-Hz band centered at $f$, would have survived Stage-0. Since the subsequent stages have a negligible false–dismissal probability ($<0.04\%$\footnote{Each of the stages contributes no more than 0.01\% false dismissal, based on the distribution of the test-signal populations.}), we take this as the upper limit consistent with the entire hierarchy of follow-ups. 

Similar to previous Einstein@Home directed searches \citep{Ming2019,Papa_2020midth,Ming2022,ming2024a,ming2025}, in each half-Hz band we  add $200$ fake continuous-wave signals, all with the same intrinsic amplitude $h_0$, into the real detector data. The data with these fake signals are then processed exactly as in the original Einstein@Home search: the data are prepared with gating and line cleaning, the Stage-0 pipeline is run, and candidates are selected via the pixeling procedure. 

A fake signal is considered recovered if:
(i) results in a candidate which falls within the containment region, and
(ii) its detection statistic value is higher than the value measured at the same parameter-space point  in the absence of the fake signal, and
(iii) its detection statistic value is higher than the 5th-highest candidate value of every pixel in that half-Hz band. 
(iv) its detection statistic value is higher than the cut-off threshold $2 \Bar{\mathcal{F}}_\mathrm{thr}$ at the Stage-0, and
(v) its detection statistic value is higher than that of any non-vetoed candidate in that half-Hz band.

\begin{figure}
  \centering
    \includegraphics[width=1.1\columnwidth]{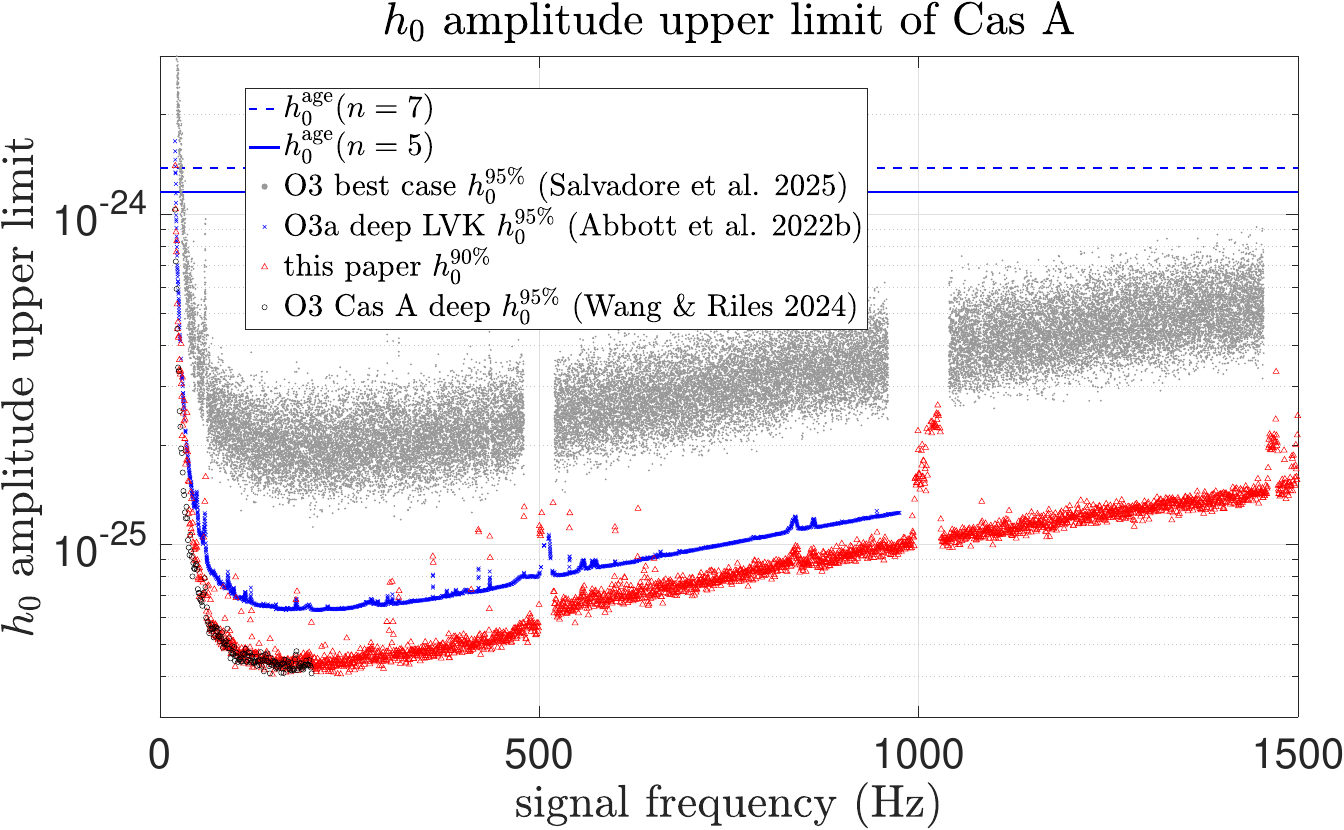}
  \caption{Upper limits on the gravitational wave amplitude of continuous gravitational wave signals from our search for Cas A (red triangles) as a function of frequency, compared to other recent results. The horizontal lines show the indirect age-based upper limits corresponding to braking indexes of 5 and 7. }
  \label{fig:UL_C}
\end{figure}

\begin{figure}
\centering
\includegraphics[width=1.1\columnwidth]{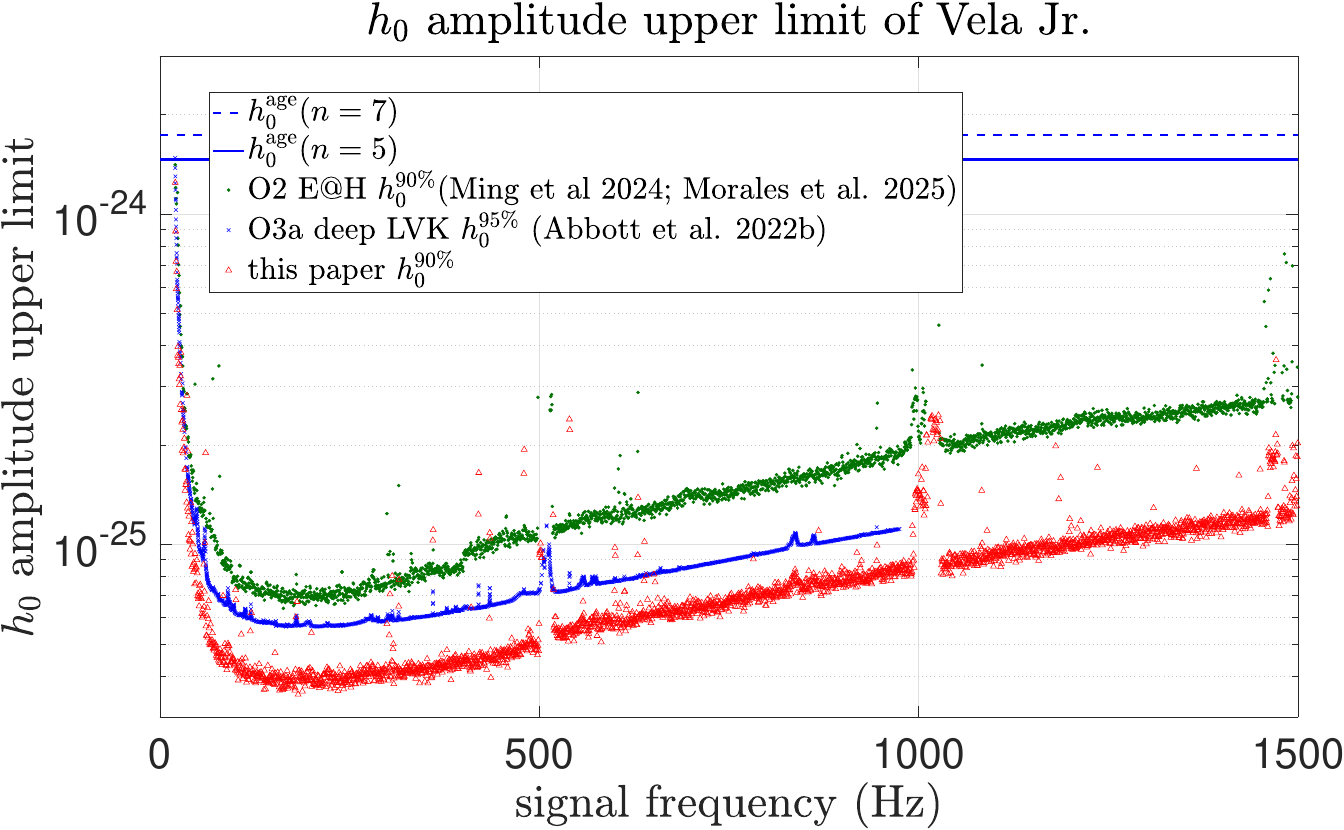}%
  \caption{Upper limits on the gravitational wave amplitude of continuous gravitational wave signals from our search for Vela Jr. (red triangles) as a function of frequency, compared to other recent results. The horizontal lines show the indirect age-based upper limits corresponding to braking indexes of 5 and 7, which are the most constraining age limit for Vela Jr., i.e.  assuming the pessimistic scenario: $(\tau=4300~\mathrm{yr}, D=750~\mathrm{pc})$. The limits under optimistic scenario $(\tau=700~\mathrm{yr}, D=200~\mathrm{pc})$ are $1.6\times 10^{-23}$ for $n=7$ and $1.4\times 10^{-23}$ for $n=5$.  }
  \label{fig:UL_V}
\end{figure}

\begin{figure}
\centering
    \includegraphics[width=1.1\columnwidth]{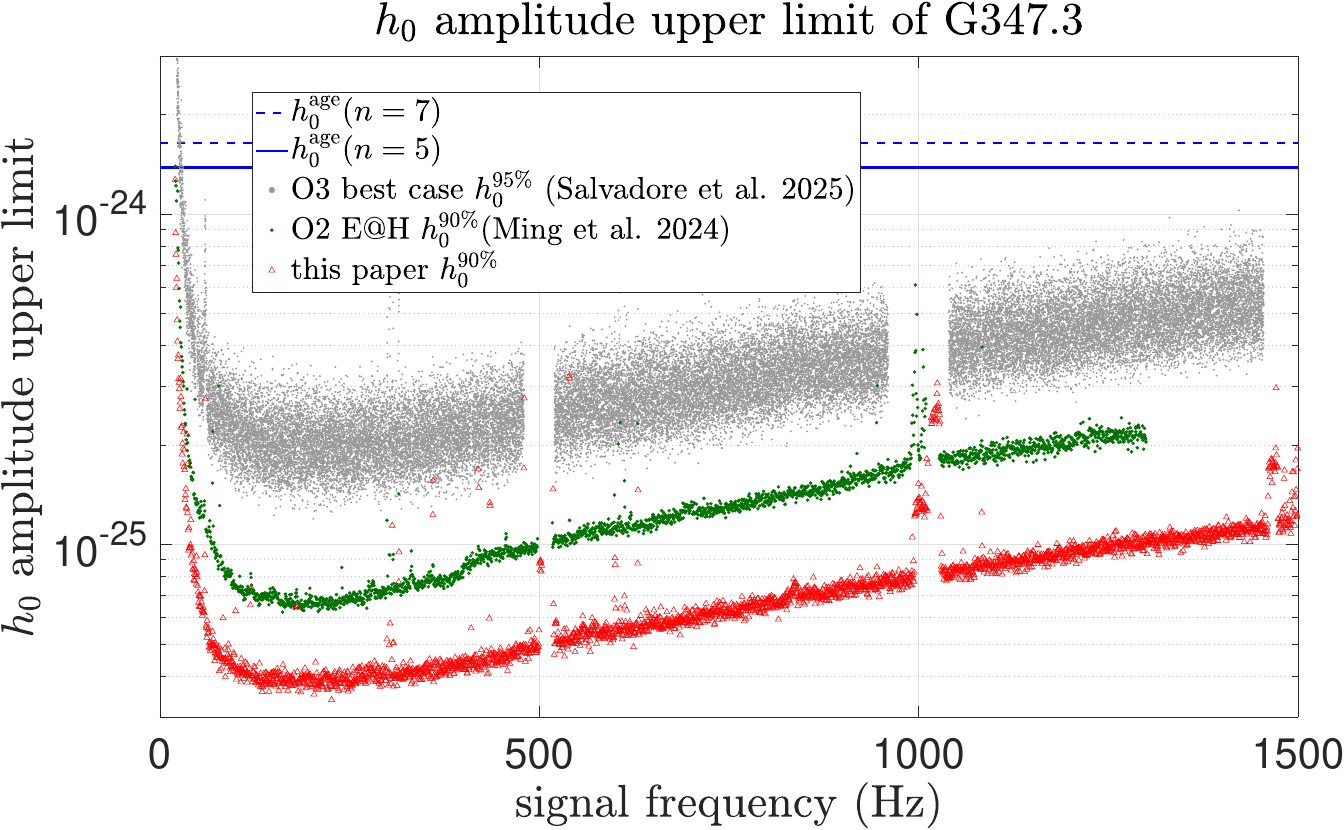}%
  \caption{Upper limits on the gravitational wave amplitude of continuous gravitational wave signals from our search for G347.3 (red triangles) as a function of frequency, compared to other recent results. The horizontal lines show the indirect age-based upper limits corresponding to braking indexes of 5 and 7. }
  \label{fig:UL_G}
\end{figure}

The fraction of recovered signals at an amplitude $h_0$ provides an estimate of the detection efficiency $C(h_0)$. Repeating this procedure for a set with 7 different $h_0$ values that bracket $C(h_0)=0.9$, we fit the data points with an error-function or logistic model 
\begin{equation}
    C(h_0) = \frac{1}{1 + \text{exp}(\tfrac{\text{a}-h_0}{\text{b}})} \ .
    \label{eqn:sigmoid_vs_h0}
\end{equation}
and use it to obtain $h_0^{90\%}(f)$, the frequentist 90\% confidence upper limits on the intrinsic GW amplitude. We apply MATLAB’s nonlinear regression method \texttt{nlpredci} to estimate the parameters $a$ and $b$ and their covariance matrix. We use this covariance to yield a $95\%$ credible interval for the fitted $h_0^{90\%}$. The overall uncertainty on the upper limit $h_0^{90\%}$ is the sum of the contribution from the efficiency–curve fit mentioned above and the instrument calibration uncertainty; for the latter we conservatively assume 5\% \citep{Cahillane2017}.
The $h_0^{90\%}$ upper limits for Cas A, Vela Jr. and G347.3 are shown in Figures~\ref{fig:UL_C}, \ref{fig:UL_V} and  \ref{fig:UL_G}, respectively.  The numerical values are provided in the Article Data as well as at \cite{AEIULurl}. 

There are some half-Hz bands where no $h_0^{90\%}$ upper limit is quoted. 
As discussed in Section~\ref{sec:data}, frequency intervals dominated by prominent spectral lines are substituted with Gaussian noise. We apply the same substitution in the search-and–recovery Monte Carlo used to determine the upper limit, after the test signals have been added. This procedure may also remove parts of the signal. Depending on how much data are substituted, the resulting detection-efficiency curve may never attain the $90\%$ level, no matter how large the $h_0$ is. In such cases, we exclude these half-Hz bands from the upper limit statements, and add them to the excluded-band list, again available in machine-readable format as the upper limit data. We exclude 7, 9, and 9 half-Hz bands between 20 and 500 Hz for Cas A, Vela Jr., and G347.3, respectively, and 42, 44, and 42 half-Hz bands between 500 and 1500 Hz for Cas A, Vela Jr., and G347.3, respectively.

In Figures~\ref{fig:UL_C}, \ref{fig:UL_V} and  \ref{fig:UL_G}, we show the resulting $h_0$ upper limits together with the most constraining upper limits from previous searches \citep{Abbott2022_4,wang2024,Salvadore2025,ming2024a,ming2025}.  
Confidence levels and upper limit procedures may differ among different searches, so the comparisons should be taken as illustrative.

{\bf{Cas A}} : our results extend the frequency range probed with O3 data by more than 500 Hz and improve in depth on previous best results \citep{Abbott2022_4}  by 42\% between 200 and 500 Hz, and by 25\% between 500 and 978 Hz.  Above 978 Hz, our results improve in depth on previous best results \citep{Salvadore2025} by 2.8 times. Below 200 Hz our results are 
comparable to those of \cite{wang2024}.  
At 200 Hz we constrain emission amplitudes to be at least 27 times smaller than what would be necessary to sustain gravitar emission throughout the life of the neutron star -- this is the age-based upper limit. See Figure~\ref{fig:UL_C}.

{\bf{Vela Jr.}}:  up to 978 Hz the most stringent previous results are from the LVK's O3a deep search \citep{Abbott2022_4}. Our results improve on these by about $37\%$. Between 978 Hz and 1500 Hz the results presented here beat the previous most sensitive ones \citep{ming2025} by about 121\%, but between 1500 Hz and 1700 Hz \citep{ming2025} remain the most sensitive.  
At 200 Hz we constrain emission amplitudes to be at least 39 times smaller than what would be necessary to sustain gravitar emission throughout the life of the neutron star, assuming the pessimistic scenario: $(\tau=4300~\mathrm{yr}, D=750~\mathrm{pc})$. Under the optimistic scenario $(\tau=700~\mathrm{yr}, D=200~\mathrm{pc})$, at 200 Hz we constrain emission amplitudes to be at least 360 times smaller than the age-based limit. See Figure~\ref{fig:UL_V}.

{\bf{G347.3}}: up to 1300 Hz, the most stringent previous results are our O2 E@H search \citep{ming2024a}. This search result  improves on these by about $93\%$. Above 1300 Hz, our results beat the previous most sensitive results \citep{Salvadore2025} by about 4.5 times. At 200 Hz we constrain emission amplitudes to be at least 39 times smaller than the age-based limit.
See Figure~\ref{fig:UL_G}.

\subsubsection{Sensitivity Depth}

For each target and every half-Hz band, we compute the search \emph{sensitivity depth} ${{\mathcal{D}}}^{90\%}$, which quantifies the sensitivity of the search itself and is approximately independent of the data sensitivity \citep{Behnke:2014tma,Dreissigacker2018}:
\begin{equation}
{{\mathcal{D}}}^{90\%}:={\sqrt{S_h(f)}\over {h_0^{90\%}(f) }}~~[ {1/\sqrt{\text{Hz}}} ],
\label{eq:sensDepth}
\end{equation}
where $\sqrt{S_h(f)}$ is the detector noise level at frequency $f$ and is computed by taking the harmonic mean over time across the entire O3a duration.

For the searches presented here the average values across the frequency ranges are:
\begin{equation}
\label{eq:sensDepthResults}
\begin{cases}
\textrm{Cas A~~~~~\,20-500 Hz}: ~&{{\mathcal{D}}}^{90\%}\approx 114~[{1/\sqrt{\text{Hz}}} ] \\
\textrm{Cas A~~~~~\,500-1500 Hz}: ~&{{\mathcal{D}}}^{90\%}\approx 91~~\,[{1/\sqrt{\text{Hz}}} ] \\
\textrm{Vela Jr.~~ 20-500 Hz}: ~&{{\mathcal{D}}}^{90\%}\approx 127~[{1/\sqrt{\text{Hz}}} ] \\
\textrm{Vela Jr.~~ 500-1500 Hz}: ~&{{\mathcal{D}}}^{90\%}\approx 108~[{1/\sqrt{\text{Hz}}} ] \\
\textrm{G347.3~~~~\,20-500 Hz}: ~&{{\mathcal{D}}}^{90\%}\approx 131~[{1/\sqrt{\text{Hz}}} ] \\
\textrm{G347.3~~~~\,500-1500 Hz}: ~&{{\mathcal{D}}}^{90\%}\approx 115~[{1/\sqrt{\text{Hz}}} ] 
\end{cases}
\end{equation}

Although the grid mismatch and $T_\mathrm{coh}$ we used for Cas A at low frequencies and Vela Jr. at high frequencies are the same, the sensitivity depth of the Cas A search is 6\% higher than that of the Vela Jr. search.
The improvement is due to the fact that for Cas A at low frequencies, we followed up a larger fraction of candidates relative to the total search templates: The high-frequency Vela Jr. template bank has roughly 3 times more templates than the low-frequency Cas A search, yet only about 2 times as many candidates were followed up. 
 This highlights that the achieved sensitivity depth depends not only on the initial search configuration (Table\ref{tab:search_params_FU}) but also on post-processing choices—most notably, how many top-ranked candidates are followed-up.

The searches presented here have sensitivity depths $10\%$ to $20\%$ higher compared to our previous Einstein@Home searches \citep{ming2024a,ming2025}, even though these use more data in the first stage. This is due to a combination of employing a longer coherent baseline at Stage-0 and following-up more candidates.

\subsection{Constraints on Astrophysical Parameters}
\label{sec:ULsRecast}

\begin{figure}[]
\centering
\subfigure{
     \includegraphics[width=1.05\columnwidth]{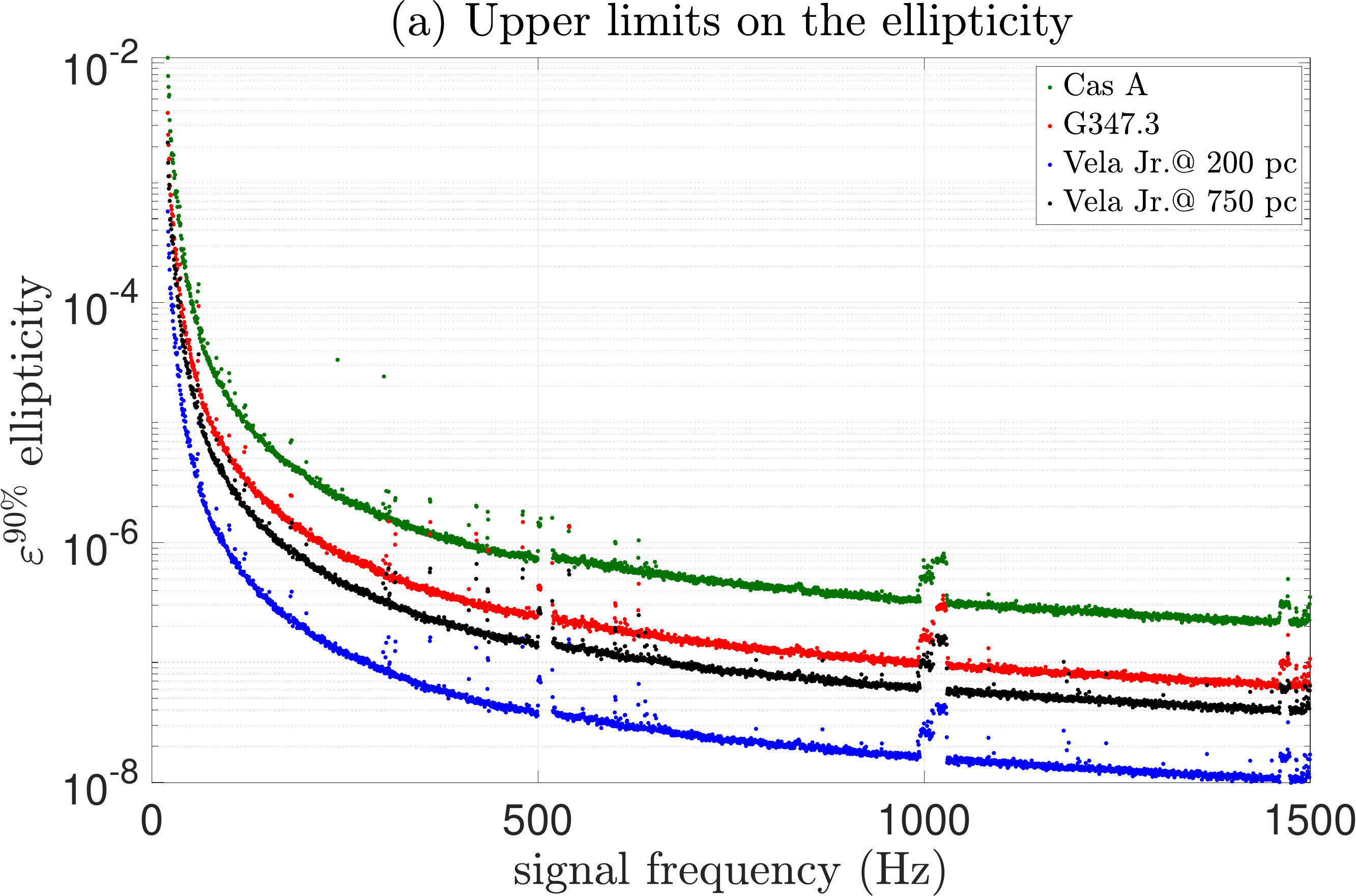}
     \label{fig:UL_e}
}\\
    \subfigure{
   \includegraphics[width=1.05\columnwidth]{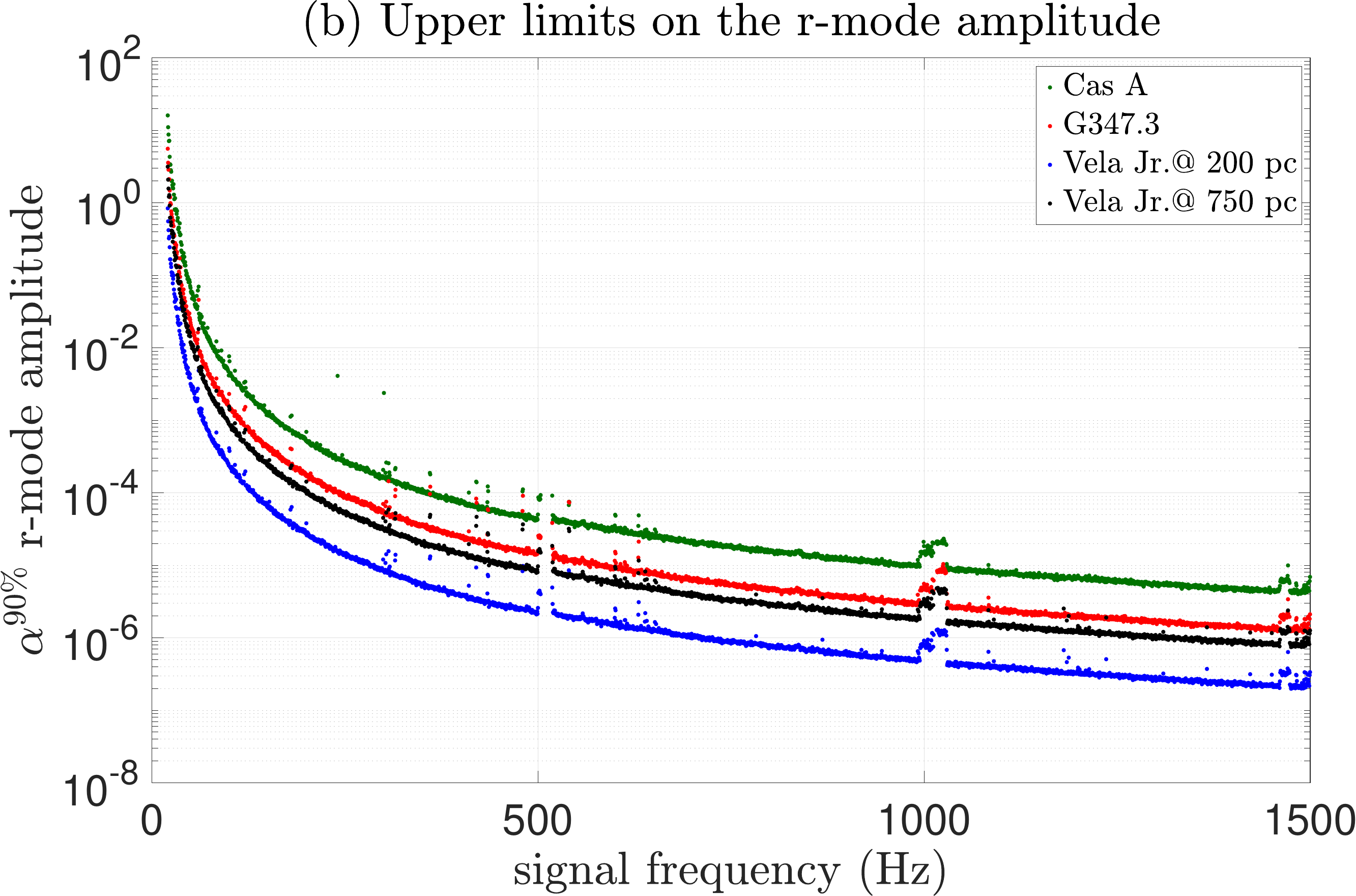}
     \label{fig:UL_a}
}
\caption{90 \% upper limits on the ellipticity (a) and r-mode amplitude (b) of the three targets as the function of gravitational wave signal frequency $f$ (twice of the spin frequency $\nu$).
For Vela Jr. we show two curves, corresponding to two distance estimates: 200 pc and 750 pc.  For Cas A, we assume 3.4 kpc.  \label{fig:ULsRecast}}
\end{figure}

The equatorial ellipticity required for a triaxial rotator, e.g. a neutron star,  to yield a continuous-wave amplitude $h_0$ at frequency $f$ is \citep{zimmermann:1979,JKS1998}
\begin{equation}
    \varepsilon = \frac{c^4}{4 \pi ^2G} \frac{h_0 D}{I f^2} \ ,
    \label{eqn:ellipticity_UL}
\end{equation}
where $c$ is the speed of light, $G$ the gravitational constant, $I$ the moment of inertia about the spin axis of the neutron star (here fixed to the conventional $I=10^{38}\,\mathrm{kg\,m^{2}}$), and $D$ the star distance. 

We can therefore recast the amplitude upper limits $h_0^{90\%}(f)$ as constraints on the equatorial ellipticity of the neutron star using Eq.~(\ref{eqn:ellipticity_UL}). 
The corresponding curves are shown in Figure~\ref{fig:UL_e}. It is expected that neutron star crusts can sustain ellipticities as large as $10^{-5}$ \citep{Ushomirsky2000,McDanielJohnsonOwen,Baiko:2018jax,Morales2022}, and young neutron stars may well present larger ellipticities than older neutron stars, acquired during the highly non-axisymmetric supernova event that generated them. Our searches probe a plausible range of ellipticities: smaller than $10^{-6}$ for all of the investigated remnants, for spin frequencies above 200 Hz. Particularly notable is the constraint for Vela Jr. at 200 pc of $\varepsilon \lesssim 4\times 10^{-8}$.

Additionally, since our search range covers frequency derivatives with braking index $n=7$, we recast the strain upper limits as constraints on the neutron star's $r$-mode oscillation amplitude, as shown in Figure \ref{fig:UL_a}.  For a canonical neutron star with mass of $1.4\,M_\odot$ and radius of 11.7 km, and moment of inertia $I=10^{38}~\mathrm{kg\,m^2}$ (as above), the dimensionless $r$-mode amplitude $\alpha$ that would yield a strain $h_0$ at GW frequency $f$ from a source at distance $D$ is \citep{Owen2010}:
\begin{equation}
	\alpha = 0.028\left( \frac{h_0}{10^{-24}} \right) \left( \frac{D}{1 \  \text{kpc}} \right) \left( \frac{f}{100 \ \text{Hz}} \right)^3 .
	\label{eqn:r_mode_amp}
\end{equation}
Young neutron stars like our targets have long been identified as good candidates for r-mode gravitational-wave emission, as they cool down after birth. Our upper limits for all targets exclude $\alpha \gtrsim 4 \times 10^{-5}$ for spin frequencies larger than 250 Hz, and for Vela Jr. at 200 pc, we can constrain $\alpha \lesssim 2 \times 10^{-6}$. With predictions for the saturation amplitudes in the range $10^{-5}-10^{-3}$ \citep{PhysRevD.76.064019,Haskell:2015iia}, these upper limits are physically interesting.

\begin{figure*}
  \centering
  \makebox[\textwidth][c]{%
    \includegraphics[width=1.0\textwidth]{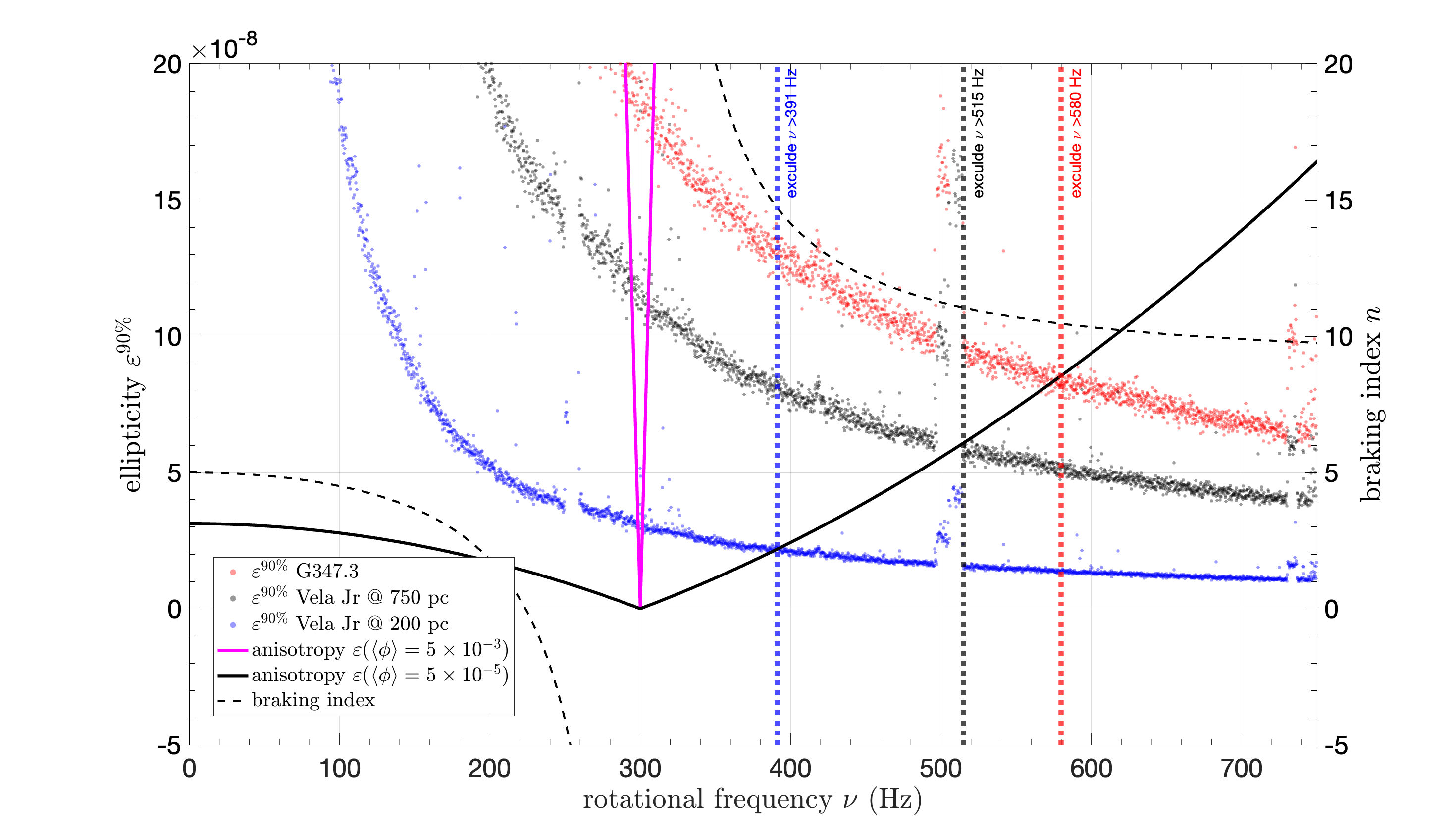}%
    }
  \caption{Constraints on the neutron-star spin frequency/crustal anisotropy. All frequencies to the right of the vertical dashed lines---up to 1500 Hz---are excluded by our search. The dashed curves denote the frequency-dependent braking index $n$, whose values are shown on the right-hand side Y axis.
 All the other curves and points' values are read from the left-hand side Y axis. }
  \label{fig:UL_ansio}
\end{figure*}

If a neutron star’s crust is even slightly anisotropic (direction-dependent stiffness), a change in spin can  turn an axisymmetric centrifugal stress into a tiny, non-axisymmetric mountain or ellipticity \citep{Morales2024}:
\begin{equation}
\varepsilon \approx \frac{m_{cr}}{M} \langle\phi\rangle \frac{\nu^2-\nu_0^2}{\nu_K^2},
\label{eq.epsilonfinal}
\end{equation}
 where $\nu$ is the rotational frequency of the neutron star, $\nu_0$ the initial rotational frequency at star's birth when the crust froze, and $\nu_K$ is the Keplerian breakup rotational frequency. $m_{cr}$ is the mass of the crust, $M$ is the total mass of the neutron star, and  $\langle\phi\rangle$ is the degree of anisotropy of the neutron star's crust.
The values of  $m_{cr}/M$ and $\nu_K$ depend on the neutron star's equation of state.


The anisotropy $\langle\phi\rangle$ is unknown, since there are no direct observational constraints for neutron-star crusts. Some theoretical models suggest that the local anisotropy in individual
Coulomb–crystal or nuclear–pasta domains can be $\gtrsim \mathcal{O}(0.1)$\citep{Baiko:2018jax,Caplan:2018gkr,Pethick:2020aey}, but when averaged over many domains the macroscopic anisotropy is expected to be much smaller because the individual domains have random orientations \citep{Blaschke:2017,Morales2024}. \cite{Morales2024} use seismic measurements of the Earth’s innermost  core to infer an average anisotropy of $\langle\phi\rangle \sim 5\times10^{-5}$, which they then adopt as a fiducial value to estimate the ellipticity of a neutron star whose crust has a comparable degree of anisotropy.

In this work, we use direct observational constraints on neutron-star ellipticity to constrain $\langle\phi\rangle$ for the first time.
We assume the conventional keplerian breakup rotational frequency $\nu_K = 1200~\mathrm{Hz}$ \citep{kelper_freq1} and the conventional crust mass ratio $m_{cr}/M=0.01$ \citep{2006Chamel}. For the  neutron star birth frequency, we assume a fiducial value $\nu_0=300$ Hz. 

In Figure~\ref{fig:UL_ansio}, the ellipticity of the neutron star due to crustal anisotropy is shown as a function of the rotational frequency $\nu$ (half of the gravitational wave frequency $\nu = f/2$) for $\langle\phi\rangle = 5\times10^{-3}$ and $\langle\phi\rangle = 5\times10^{-5}$. At $\nu= 300~\mathrm{Hz}$ the anisotropy–induced ellipticity vanishes, indicating that this frequency corresponds to the initial rotational frequency at crust formation, $\nu_0 = 300~\mathrm{Hz}$.
Our observational upper limits on the ellipticity are superimposed on this plot and identify the spin frequencies excluded by our results. Assuming $\langle\phi\rangle = 5\times10^{-5}$, we can exclude spin frequencies above 580 Hz,  515 Hz, and 391 Hz for G347.3, Vela Jr. at 750 pc, and Vela Jr. at 200 pc, respectively. In the case of $\langle\phi\rangle = 5\times10^{-3}$, we can exclude nearly all spin periods between 1.3--100 ms. Conversely, we can conclude that unless the neutron stars in our targets spin slower than 10 Hz or higher than 750 Hz,  it is unlikely that their crustal anisotropy is higher than  $5\times10^{-3}$.

\begin{figure*}[]
   \includegraphics[width=0.95\textwidth]{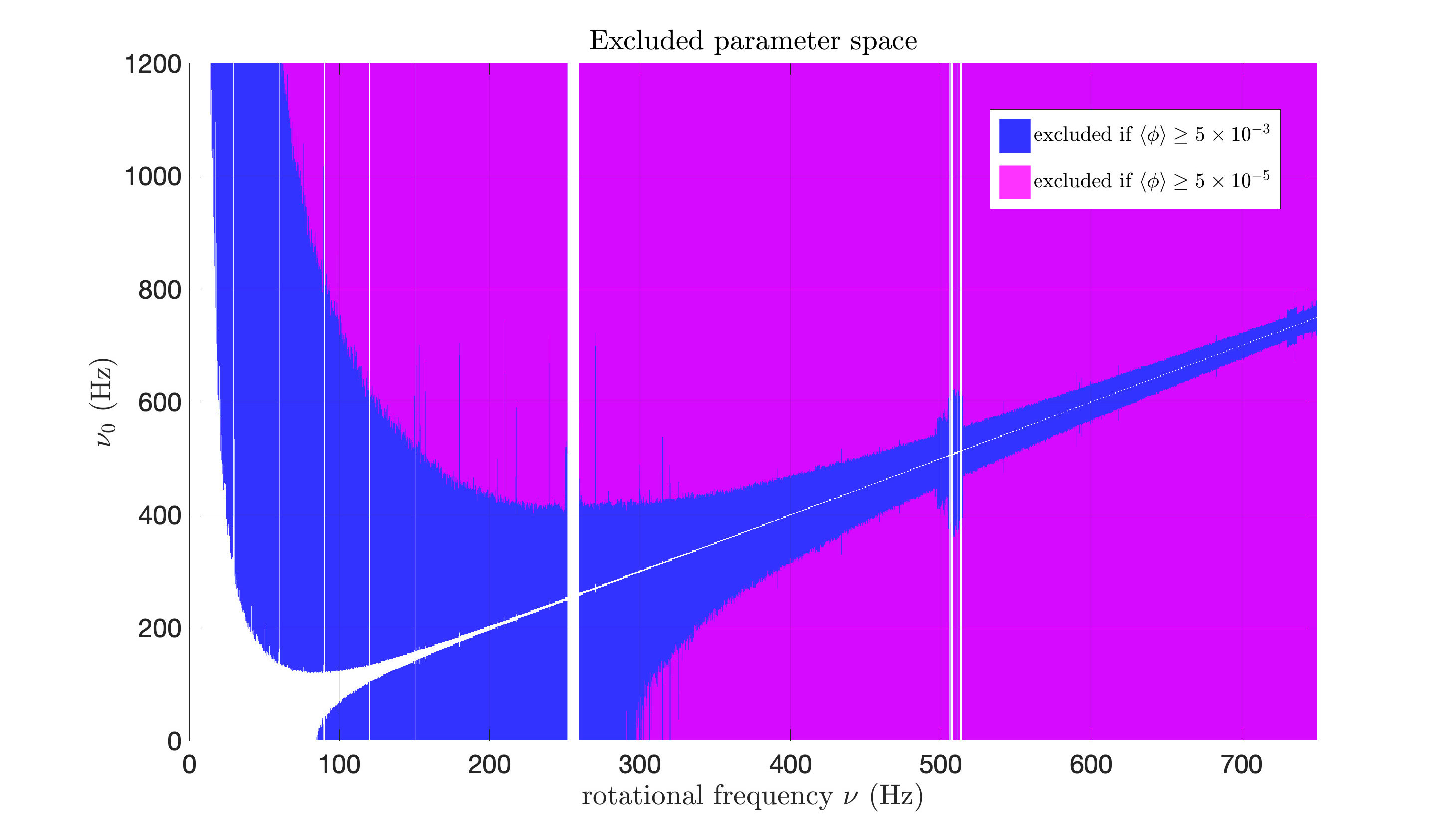}
\caption{Excluded regions in the $(\nu_0,\nu)$ plane inferred from our most constraining ellipticity upper limits for Vela Jr.\ assuming $d=200$ pc. Shaded areas indicate combinations of birth spin frequency $\nu_0$ and current spin frequency $\nu$ that are ruled out; blue and magenta correspond to $\langle\phi\rangle=5\times10^{-3}$ and $\langle\phi\rangle=5\times10^{-5}$, respectively (overlap indicates exclusion for both values of $\langle\phi\rangle$). Vertical white gaps denote frequency bands where no reliable upper limits could be set (e.g., contaminated frequency bands of O3a data), and thus no exclusion can be claimed there.}
\label{fig:anisotropy_2}
\end{figure*}

Since the anisotropy–induced ellipticity is frequency dependent, the corresponding braking
index $n$ is also frequency dependent. Starting from the definition of the
braking index, $n = \frac{\nu \ddot{\nu}}{\dot{\nu}^2}$, and combining
this with Eq.~(\ref{eq.epsilonfinal}), we obtain
\begin{equation}
n(\nu) = 5 + \frac{4 \nu^{2}}{\nu^{2} - \nu_0^{2}} \, .
\label{eq:n_of_f_explicit}
\end{equation}
The braking index $n(\nu)$ is shown on the right-hand $y$-axis of Figure~\ref{fig:UL_ansio}. In the excluded frequency range, the effective braking index can reach values as high as 14.7 (blue vertical dashed line), which nevertheless remain within our searched parameter space because, as previously explained, the parameter space bounds of Eq.~(\ref{eqn:param_space}) are safely inclusive for all $n \geq 2$.

The birth spin frequency of a neutron star $\nu_0$ can in principle be any number below its Keplerian breakup frequency $\nu_K$ (see for example the discussion in \cite{Pagliaro:2023bvi}). 
Stellar evolution and collapse calculations, as well as magnetar/GRB central-engine scenarios, suggest that a subset of newly born neutron stars may be born with very fast spin $300~\mathrm{Hz} \lesssim \nu_0 \lesssim 1200~\mathrm{Hz}$ \citep{Heger:2003nh,Ott:2005wh,Metzger2011}. However, independent observational inferences for ordinary radio pulsars favor substantially slower birth spins. Using pulsars associated with supernova remnants and assuming magneto-dipole spin-down, initial periods are typically inferred to be tens–hundreds of milliseconds \citep{PopovTurolla2012}.
More recently, a hierarchical Bayesian analysis of pulsars associated with supernova remnants finds an initial-period distribution peaking at $\sim$ 50 ms \citep{Du:2024gfi}. 
Population-synthesis studies support birth periods extending to several hundred milliseconds \citep{Faucher-Giguere:2005dxp}.

Given the large theoretical and observational uncertainty in neutron-star birth spins, it is instructive to treat the initial spin frequency $\nu_0$ as a free parameter and explore how our constraints vary across the $(\nu_0, \nu)$ plane. In particular, we map parameter space spanned by $\nu_0$ and the spin frequency $\nu$ using our most stringent ellipticity upper limits, which are obtained from the Vela Jr. target under the optimistic distance assumption of 200 pc. 
This choice provides the strongest leverage on $\langle\phi\rangle$ and allows us to illustrate how direct observational upper limits on $\varepsilon$ translate into constraints on the underlying model parameters across the full range of plausible birth spins, from conservative values $(\nu_0 \sim$ a few Hz) up to rapid-spin scenarios $\nu_0 \lesssim \nu_K$.

In Figure~\ref{fig:anisotropy_2} we show the $(\nu_0,\nu)$ regions excluded by our search results, under different assumptions for the anisotropy $\langle\phi\rangle$.  
When $\langle\phi\rangle$ is sufficiently large (e.g., $\langle\phi\rangle = 5\times10^{-3}$), our ellipticity upper limits exclude most of the investigated spin frequency range, leaving only a small fraction of the $(\nu_0,\nu)$ plane unconstrained.
For smaller anisotropy values (e.g., $\langle\phi\rangle = 5\times10^{-5}$), the excluded region becomes more structured and the results are more sensitive to the assumed birth spin $\nu_0$. The excluded fraction of the searched frequency band decreases from $59\%$ at $\nu_0=2~\mathrm{Hz}$ to a minimum of $37\%$ at $\nu_0=410~\mathrm{Hz}$, and then increases again, reaching $90\%$ at $\nu_0=1200~\mathrm{Hz}$.

\section{Acknowledgments}

We extend our heartfelt thanks to the many thousands of Einstein@Home volunteers whose donated CPU/GPU time made this work possible. 
We also thank the LIGO instrument scientists and engineers, whose remarkable efforts have delivered detectors sensitive to gravitational-wave strains at astonishingly small levels. We acknowledge the LIGO data calibration and line identification experts whose work helps us prepare the input data to our searches.
Substantial post-processing is performed on the ATLAS cluster at AEI Hannover; we are grateful to Bruce Allen, Carsten Aulbert and Henning Fehrmann for their steadfast support. 
This research has made use of data and web tools for data download obtained from the Gravitational Wave Open Science Center (https://www.gw-openscience.org/), a service of LIGO Laboratory, the LIGO Scientific Collaboration and the Virgo Collaboration. 
%

\appendix
\section{G347.3 Parameters}
\begin{figure*}[t]
\centering
\includegraphics[width=0.5\textwidth]{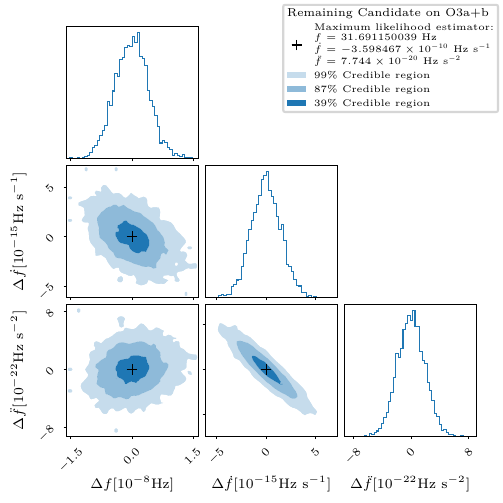}
\caption{O3a+b Bayesian follow-up posteriors and credible intervals around the maximum likelihood estimator.}
\label{fig:posteriorG347cand}
\end{figure*}

The best estimate for the G347.3 surviving candidate parameters is given in Table~\ref{tab:G347cand}, from the posteriors of Figure~\ref{fig:posteriorG347cand}. These parameters are based on the longest stretch of contiguous data available to us -- O3a+b -- and leverage the Bayesian follow-up method of \cite{Martins:2025jnq}. 

\FloatBarrier

\begin{table*}[t]
\centering
\begin{tabular}{| c | c | c | c | c | c |}
\hline\hline
$f$ [Hz] & $\dot{f}$ [Hz/s] & $\ddot{f}$ [Hz/s$^2$] &
$\Delta^{99\%} f$ [Hz] & $\Delta^{99\%} \dot{f}$ [Hz/s] & $\Delta^{99\%} \ddot{f}$ [Hz/s$^2$] \\
\hline
$31.691150039$ & $-3.598467\times 10^{-10}$ & $7.744\times 10^{-20}$ &
$\pm 1.5\times 10^{-8}$ & $\pm 5.0 \times 10^{-15}$ & $\pm 8.0 \times 10^{-22}$ \\
\hline\hline
\end{tabular}
\caption{\label{tab:G347cand} Maximum likelihood estimator values and 99\% credible intervals from the O3a+b follow-up of the surviving G347.3 candidate.}
\end{table*}

%
%
%

\bibliography{bibliography}

@article{LIGOScientific:2025slb,
    author = "Abac, A. G. and others",
    collaboration = "LIGO Scientific, VIRGO, KAGRA",
    title = "{GWTC-4.0: Updating the Gravitational-Wave Transient Catalog with Observations from the First Part of the Fourth LIGO-Virgo-KAGRA Observing Run}",
    eprint = "2508.18082",
    archivePrefix = "arXiv",
    primaryClass = "gr-qc",
    reportNumber = "LIGO-P2400386",
    month = "8",
    year = "2025"
}

@article{Dergachev:2024knd,
    author = "Dergachev, Vladimir and Papa, Maria Alessandra",
    title = "{Early release of the expanded atlas of the sky in continuous gravitational waves}",
    eprint = "2401.13173",
    archivePrefix = "arXiv",
    primaryClass = "gr-qc",
    doi = "10.1103/PhysRevD.109.022007",
    journal = "Phys. Rev. D",
    volume = "109",
    number = "2",
    pages = "022007",
    year = "2024"
}

@article{Jones:2024nty,
    author = "Jones, D. I. and Hutchins, T. J.",
    title = "{Crustal lattice pressure as a source of neutron star mountains}",
    eprint = "2407.00162",
    archivePrefix = "arXiv",
    primaryClass = "astro-ph.HE",
    doi = "10.1093/mnras/staf784",
    journal = "Mon. Not. Roy. Astron. Soc.",
    volume = "540",
    number = "3",
    pages = "2349--2358",
    year = "2025"
}

@article{LIGOScientific:2008hqb,
    author = "Wette, K. and others",
    editor = "Scott, Susan M. and McClelland, David E.",
    collaboration = "LIGO Scientific",
    title = "{Searching for gravitational waves from Cassiopeia A with LIGO}",
    eprint = "0802.3332",
    archivePrefix = "arXiv",
    primaryClass = "gr-qc",
    reportNumber = "LIGO-P070123-01-Z",
    doi = "10.1088/0264-9381/25/23/235011",
    journal = "Class. Quant. Grav.",
    volume = "25",
    pages = "235011",
    year = "2008"
}

@article{Krishnan:2004sv,
    author = "Krishnan, Badri and Sintes, Alicia M. and Papa, Maria Alessandra and Schutz, Bernard F. and Frasca, Sergio and Palomba, Cristiano",
    title = "{The Hough transform search for continuous gravitational waves}",
    eprint = "gr-qc/0407001",
    archivePrefix = "arXiv",
    reportNumber = "AEI-2004-050",
    doi = "10.1103/PhysRevD.70.082001",
    journal = "Phys. Rev. D",
    volume = "70",
    pages = "082001",
    year = "2004"
}

@article{Riles:2022wwz,
    author = "Riles, Keith",
    title = "{Searches for continuous-wave gravitational radiation}",
    eprint = "2206.06447",
    archivePrefix = "arXiv",
    primaryClass = "astro-ph.HE",
    doi = "10.1007/s41114-023-00044-3",
    journal = "Living Rev. Rel.",
    volume = "26",
    number = "1",
    pages = "3",
    year = "2023"
}

@article{Du:2024gfi,
    author = "Du, Shen-Shi and Liu, Xiao-Jin and Chen, Zu-Cheng and You, Zhi-Qiang and Zhu, Xing-Jiang and Zhu, Zong-Hong",
    title = "{On the Initial Spin Period Distribution of Neutron Stars}",
    eprint = "2402.14030",
    archivePrefix = "arXiv",
    primaryClass = "astro-ph.HE",
    doi = "10.3847/1538-4357/ad4450",
    journal = "Astrophys. J.",
    volume = "968",
    number = "2",
    pages = "105",
    year = "2024"
}

@article{Pagliaro:2023bvi,
    author = "Pagliaro, Gianluca and Papa, Maria Alessandra and Ming, Jing and Lian, Jianhui and Tsuna, Daichi and Maraston, Claudia and Thomas, Daniel",
    title = "{Continuous Gravitational Waves from Galactic Neutron Stars: Demography, Detectability, and Prospects}",
    eprint = "2303.04714",
    archivePrefix = "arXiv",
    primaryClass = "gr-qc",
    doi = "10.3847/1538-4357/acd76f",
    journal = "Astrophys. J.",
    volume = "952",
    number = "2",
    pages = "123",
    year = "2023"
}

@article{Faucher-Giguere:2005dxp,
    author = "Faucher-Giguere, Claude-Andre and Kaspi, Victoria M.",
    title = "{Birth and evolution of isolated radio pulsars}",
    eprint = "astro-ph/0512585",
    archivePrefix = "arXiv",
    doi = "10.1086/501516",
    journal = "Astrophys. J.",
    volume = "643",
    pages = "332--355",
    year = "2006"
}

@ARTICLE{PopovTurolla2012,
       author = {{Popov}, S.~B. and {Turolla}, R.},
        title = "{Initial spin periods of neutron stars in supernova remnants}",
      journal = {\apss},
         year = 2012,
        month = oct,
       volume = {341},
       number = {2},
        pages = {457-464},
          doi = {10.1007/s10509-012-1100-z},
archivePrefix = {arXiv},
       eprint = {1204.0632},
 primaryClass = {astro-ph.HE},
       adsurl = {https://ui.adsabs.harvard.edu/abs/2012Ap&SS.341..457P}
}

@article{Green:2024uci,
    author = "Green, D. A.",
    title = "{An updated catalogue of 310 Galactic supernova remnants and their statistical properties}",
    eprint = "2411.03367",
    archivePrefix = "arXiv",
    primaryClass = "astro-ph.GA",
    doi = "10.1007/s12036-024-10038-4",
    journal = "J. Astrophys. Astron.",
    volume = "46",
    number = "1",
    pages = "14",
    year = "2025"
}

@misc{Green_2024,
author    = {{Green, D. A.}},
  title     = {A Catalogue of Galactic Supernova Remnants, October 2024 version},
  year      = {2026},
  publisher = {Cavendish Laboratory, Cambridge, United Kingdom},
  url       = {https://www.mrao.cam.ac.uk/surveys/snrs/},
  version   = {Last Accessed: 2026-03-18}
}

@article{Leike:2020jyl,
    author = "Leike, Reimar and Celli, Silvia and Krone-Martins, Alberto and Boehm, Celine and Glatzle, Martin and Fukui, Yasou and Sano, Hidetoshi and Rowell, Gavin",
    title = "{Optical reconstruction of dust in the region of supernova remnant RX J1713.7{\ensuremath{-}}3946 from astrometric data}",
    eprint = "2011.14383",
    archivePrefix = "arXiv",
    primaryClass = "astro-ph.HE",
    doi = "10.1038/s41550-021-01344-w",
    journal = "Nature Astron.",
    volume = "5",
    number = "8",
    pages = "832--838",
    year = "2021"
}

@article{Martins:2025jnq,
    author = "Martins, Jasper and Papa, Maria Alessandra and Steltner, Benjamin and Prix, Reinhard and Vidal, P. B.",
    title = "{A Bayesian Framework to Follow-up Continuous Gravitational Wave Candidates from Deep Surveys}",
    eprint = "2508.18204",
    archivePrefix = "arXiv",
    primaryClass = "gr-qc",
    month = "8",
    year = "2025"
}

@ARTICLE{brian2025b,
       author = {{McGloughlin}, Brian and {Steltner}, Benjamin and {Martins}, Jasper and {Alessandra Papa}, Maria and {Eggenstein}, Heinz-Bernd and {Ming}, Jing and {Machenschalk}, Bernd and {Prix}, Reinhard and {Bensch}, Maximillian},
        title = "{High-frequency continuous gravitational waves searched in LIGO O3 public data with Einstein@Home}",
      journal = {arXiv e-prints},
         year = 2025,
        month = aug,
          eid = {arXiv:2508.20073},
        pages = {arXiv:2508.20073},
          doi = {10.48550/arXiv.2508.20073},
archivePrefix = {arXiv},
       eprint = {2508.20073},
 primaryClass = {gr-qc},
       adsurl = {https://ui.adsabs.harvard.edu/abs/2025arXiv250820073M}
}

@ARTICLE{brian2025a,
       author = {{McGloughlin}, Brian and {Martins}, Jasper and {Steltner}, Benjamin and {Alessandra Papa}, Maria and {Eggenstein}, Heinz-Bernd and {Machenschalk}, Bernd and {Prix}, Reinhard and {Bensch}, Maximillian},
        title = "{Einstein@Home all-sky ``bucket'' search for continuous gravitational waves in LIGO O3 public data}",
      journal = {arXiv e-prints},
         year = 2025,
        month = aug,
          eid = {arXiv:2508.16423},
        pages = {arXiv:2508.16423},
          doi = {10.48550/arXiv.2508.16423},
archivePrefix = {arXiv},
       eprint = {2508.16423},
 primaryClass = {gr-qc},
       adsurl = {https://ui.adsabs.harvard.edu/abs/2025arXiv250816423M}
}

@ARTICLE{GW150914,
       author = {{Abbott}, B.~P. and others},
        title = "{GW150914: First results from the search for binary black hole coalescence with Advanced LIGO}",
      journal = {\prd},
         year = 2016,
        month = jun,
       volume = {93},
       number = {12},
          eid = {122003},
        pages = {122003},
          doi = {10.1103/PhysRevD.93.122003},
archivePrefix = {arXiv},
       eprint = {1602.03839},
 primaryClass = {gr-qc},
       adsurl = {https://ui.adsabs.harvard.edu/abs/2016PhRvD..93l2003A}
}

@ARTICLE{Salvadore2025,
  title = {Harnessing the potential of pystoch: Detecting continuous gravitational waves from interesting supernova remnant targets},
  author = {Salvadore, Claudio and La Rosa, Iuri and Leaci, Paola and Amicucci, Francesco and Astone, Pia and D'Antonio, Sabrina and D'Onofrio, Luca and Palomba, Cristiano and Pierini, Lorenzo and Tehrani, Francesco Safai},
  journal = {Phys. Rev. D},
  volume = {112},
  issue = {8},
  pages = {083051},
  numpages = {10},
  year = {2025},
  month = {Oct},
  publisher = {American Physical Society},
  doi = {10.1103/xydb-k4vw},
  url = {https://link.aps.org/doi/10.1103/xydb-k4vw}
}

@ARTICLE{2008A&A...484..457M,
   author = {{Mignani}, R.~P. and {Zaggia}, S. and {de Luca}, A. and {Perna}, R. and 
 {Bassan}, N. and {Caraveo}, P.~A.},
    title = "{Optical and infrared observations of the X-ray source 1WGA J1713.4-3949 in the G347.3-0.5 SNR}",
  journal = {\aap},
archivePrefix = "arXiv",
   eprint = {0803.3722},
     year = 2008,
    month = jun,
   volume = 484,
    pages = {457-461},
      doi = {10.1051/0004-6361:20079076},
   adsurl = {http://adsabs.harvard.edu/abs/2008A%26A...484..457M}
}

@ARTICLE{2004A&A...427..199C,
   author = {{Cassam-Chena{\"i}}, G. and {Decourchelle}, A. and {Ballet}, J. and 
 {Sauvageot}, J.-L. and {Dubner}, G. and {Giacani}, E.},
    title = "{XMM-Newton observations of the supernova remnant RX J1713.7-3946 and its central source}",
  journal = {\aap},
   eprint = {astro-ph/0407333},
     year = 2004,
    month = nov,
   volume = 427,
    pages = {199-216},
      doi = {10.1051/0004-6361:20041154},
   adsurl = {http://adsabs.harvard.edu/abs/2004A%26A...427..199C}
}

@ARTICLE{1997A&A...318L..59W,
   author = {{Wang}, Z.~R. and {Qu}, Q.-Y. and {Chen}, Y.},
    title = "{Is RX J1713.7-3946 the remnant of the AD393 guest star?}",
  journal = {\aap},
     year = 1997,
    month = feb,
   volume = 318,
    pages = {L59-L61},
   adsurl = {http://adsabs.harvard.edu/abs/1997A%26A...318L..59W}
}

@ARTICLE{2012AJ....143...27F,
       author = {{Fesen}, Robert A. and {Kremer}, Richard and {Patnaude}, Daniel and {Milisavljevic}, Dan},
        title = "{The SN 393-SNR RX J1713.7-3946 (G347.3-0.5) Connection}",
      journal = {\aj},
         year = 2012,
        month = feb,
       volume = {143},
       number = {2},
          eid = {27},
        pages = {27},
          doi = {10.1088/0004-6256/143/2/27},
archivePrefix = {arXiv},
       eprint = {1112.0593},
 primaryClass = {astro-ph.HE},
       adsurl = {https://ui.adsabs.harvard.edu/abs/2012AJ....143...27F}
}

@article{Cahillane2017,
  title = {Calibration uncertainty for Advanced LIGO's first and second observing runs},
  author = {Cahillane, Craig and Betzwieser, Joe and Brown, Duncan A. and Goetz, Evan and Hall, Evan D. and Izumi, Kiwamu and Kandhasamy, Shivaraj and Karki, Sudarshan and Kissel, Jeff S. and Mendell, Greg and Savage, Richard L. and Tuyenbayev, Darkhan and Urban, Alex and Viets, Aaron and Wade, Madeline and Weinstein, Alan J.},
  journal = {Phys. Rev. D},
  volume = {96},
  issue = {10},
  pages = {102001},
  numpages = {16},
  year = {2017},
  month = {Nov},
  publisher = {American Physical Society},
  doi = {10.1103/PhysRevD.96.102001},
  url = {https://link.aps.org/doi/10.1103/PhysRevD.96.102001}
}

@ARTICLE{McDanielJohnsonOwen, 
author	={N.K.~Johnson-McDaniel and B.J.~Owen},
title= "{Maximum elastic deformations of relativistic stars}",
journal={Phys. Rev. D},
volume={88},
year="2013",
pages={044004}
}

@ARTICLE{kelper_freq1,
       author = {{Haensel}, P. and {Zdunik}, J.~L. and {Bejger}, M. and {Lattimer}, J.~M.},
        title = "{Keplerian frequency of uniformly rotating neutron stars and strange stars}",
      journal = {\aap},
         year = 2009,
        month = aug,
       volume = {502},
       number = {2},
        pages = {605-610},
          doi = {10.1051/0004-6361/200811605},
archivePrefix = {arXiv},
       eprint = {0901.1268},
 primaryClass = {astro-ph.SR},
       adsurl = {https://ui.adsabs.harvard.edu/abs/2009A&A...502..605H}
}

@ARTICLE{2006Chamel,
       author = {{Chamel}, Nicolas},
        title = "{Effective mass of free neutrons in neutron star crust}",
      journal = {\nphysa},
         year = 2006,
        month = jul,
       volume = {773},
       number = {3-4},
        pages = {263-278},
          doi = {10.1016/j.nuclphysa.2006.04.010},
archivePrefix = {arXiv},
       eprint = {nucl-th/0512034},
 primaryClass = {nucl-th},
       adsurl = {https://ui.adsabs.harvard.edu/abs/2006NuPhA.773..263C}
}

@article{Pethick:2020aey,
    author = "Pethick, C. J. and Zhang, Zhaowen and Kobyakov, D. N.",
    title = "{Elastic properties of phases with nonspherical nuclei in dense matter}",
    eprint = "2003.13430",
    archivePrefix = "arXiv",
    primaryClass = "cond-mat.mtrl-sci",
    reportNumber = "NORDITA 2020-31",
    doi = "10.1103/PhysRevC.101.055802",
    journal = "Phys. Rev. C",
    volume = "101",
    number = "5",
    pages = "055802",
    year = "2020"
}

@ARTICLE{Blaschke:2017,
       author = {{Blaschke}, Daniel N.},
        title = "{Averaging of elastic constants for polycrystals}",
      journal = {Journal of Applied Physics},
         year = 2017,
        month = oct,
       volume = {122},
       number = {14},
          eid = {145110},
        pages = {145110},
          doi = {10.1063/1.4993443},
archivePrefix = {arXiv},
       eprint = {1706.07132},
 primaryClass = {cond-mat.mtrl-sci},
       adsurl = {https://ui.adsabs.harvard.edu/abs/2017JAP...122n5110B}
}

@article{Caplan:2018gkr,
    author = "Caplan, M. E. and Schneider, A. S. and Horowitz, C. J.",
    title = "{Elasticity of Nuclear Pasta}",
    eprint = "1807.02557",
    archivePrefix = "arXiv",
    primaryClass = "nucl-th",
    doi = "10.1103/PhysRevLett.121.132701",
    journal = "Phys. Rev. Lett.",
    volume = "121",
    number = "13",
    pages = "132701",
    year = "2018"
}

@article{Baiko:2018jax,
    author = "Baiko, D. A. and Chugunov, A. I.",
    title = "{Breaking properties of neutron star crust}",
    eprint = "1808.06415",
    archivePrefix = "arXiv",
    primaryClass = "astro-ph.HE",
    doi = "10.1093/mnras/sty2259",
    journal = "Mon. Not. Roy. Astron. Soc.",
    volume = "480",
    number = "4",
    pages = "5511--5516",
    year = "2018"
}

@article{Heger:2003nh,
    author = "Heger, A. and Woosley, S. E. and Langer, N. and Spruit, H. C.",
    title = "{Presupernova evolution of rotating massive stars and the rotation rate of pulsars}",
    eprint = "astro-ph/0301374",
    archivePrefix = "arXiv",
    journal = "IAU Symp.",
    volume = "215",
    pages = "591",
    year = "2004"
}

@article{Ott:2005wh,
    author = "Ott, Christian D. and Burrows, Adam and Thompson, Todd A. and Livne, Eli and Walder, Rolf",
    title = "{The spin periods and rotational profiles of neutron stars at birth}",
    eprint = "astro-ph/0508462",
    archivePrefix = "arXiv",
    doi = "10.1086/500832",
    journal = "Astrophys. J. Suppl.",
    volume = "164",
    pages = "130--155",
    year = "2006"
}

@ARTICLE{Metzger2011,
       author = {{Metzger}, B.~D. and {Giannios}, D. and {Thompson}, T.~A. and {Bucciantini}, N. and {Quataert}, E.},
        title = "{The protomagnetar model for gamma-ray bursts}",
      journal = {\mnras},
         year = 2011,
        month = may,
       volume = {413},
       number = {3},
        pages = {2031-2056},
          doi = {10.1111/j.1365-2966.2011.18280.x},
archivePrefix = {arXiv},
       eprint = {1012.0001},
 primaryClass = {astro-ph.HE},
       adsurl = {https://ui.adsabs.harvard.edu/abs/2011MNRAS.413.2031M}
}

@article{wang2024,
  title = {Deep search of the full O3 LIGO data for continuous gravitational waves from the Cassiopeia A central compact object},
  author = {Wang, Jonathan and Riles, Keith},
  journal = {Phys. Rev. D},
  volume = {110},
  issue = {4},
  pages = {042006},
  numpages = {13},
  year = {2024},
  month = {Aug},
  publisher = {American Physical Society},
  doi = {10.1103/PhysRevD.110.042006},
  url = {https://link.aps.org/doi/10.1103/PhysRevD.110.042006}
}

@article{zimmermann:1979,
  title = {Gravitational waves from rotating and precessing rigid bodies: Simple models and applications to pulsars},
  author = {Zimmermann, Mark and Szedenits, Eugene},
  journal = {Phys. Rev. D},
  volume = {20},
  issue = {2},
  pages = {351--355},
  numpages = {0},
  year = {1979},
  month = {Jul},
  publisher = {American Physical Society},
  doi = {10.1103/PhysRevD.20.351},
  url = {http://link.aps.org/doi/10.1103/PhysRevD.20.351}
}

@article{Abbott_2023,
   title={Open Data from the Third Observing Run of LIGO, Virgo, KAGRA, and GEO},
   volume={267},
   ISSN={1538-4365},
   url={http://dx.doi.org/10.3847/1538-4365/acdc9f},
   DOI={10.3847/1538-4365/acdc9f},
   number={2},
   journal={The Astrophysical Journal Supplement Series},
   publisher={American Astronomical Society},
   author = "Abbott, Rich and others",
   year={2023},
   month=jul, pages={29} }

@misc{covas2024,
      title={Search for continuous gravitational waves from unknown neutron stars in binary systems with long orbital periods in O3 data}, 
      author={P. B. Covas and M. A. Papa and R. Prix},
      year={2024},
      eprint={2409.16196},
      archivePrefix={arXiv},
      primaryClass={gr-qc},
      url={https://arxiv.org/abs/2409.16196}, 
}

@article{Dergachev:2025ead,
    author = "Dergachev, Vladimir and Papa, Maria Alessandra",
    title = "{Early release of low-frequency atlas of continuous gravitational waves}",
    eprint = "2507.12161",
    archivePrefix = "arXiv",
    primaryClass = "gr-qc",
    month = "7",
    year = "2025"
}

@article{Abbott2022_3,
  title = {All-sky search for continuous gravitational waves from isolated neutron stars using Advanced LIGO and Advanced Virgo O3 data},
  author = {Abbott, R. and others},
  journal = {Phys. Rev. D},
  volume = {106},
  issue = {10},
  pages = {102008},
  numpages = {37},
  year = {2022},
  month = {Nov},
  publisher = {American Physical Society},
  doi = {10.1103/PhysRevD.106.102008},
  url = {https://link.aps.org/doi/10.1103/PhysRevD.106.102008}
}

@article{Steltner2023,
doi = {10.3847/1538-4357/acdad4},
url = {https://dx.doi.org/10.3847/1538-4357/acdad4},
year = {2023},
month = {jul},
publisher = {The American Astronomical Society},
volume = {952},
number = {1},
pages = {55},
author = {B. Steltner and M. A. Papa and H.-B. Eggenstein and R. Prix and M. Bensch and B. Allen and B. Machenschalk},
title = {Deep Einstein@Home All-sky Search for Continuous Gravitational Waves in LIGO O3 Public Data},
journal = {The Astrophysical Journal}
}

@article{Dergachev2023,
    author = "Dergachev, Vladimir and Papa, Maria Alessandra",
    title = "{Frequency-Resolved Atlas of the Sky in Continuous Gravitational Waves}",
    eprint = "2202.10598",
    archivePrefix = "arXiv",
    primaryClass = "gr-qc",
    doi = "10.1103/PhysRevX.13.021020",
    journal = "Phys. Rev. X",
    volume = "13",
    number = "2",
    pages = "021020",
    year = "2023"
}

@article{LIGOScientific:2021mwx,
   title={Searches for Continuous Gravitational Waves from Young Supernova Remnants in the Early Third Observing Run of Advanced LIGO and Virgo},
   volume={921},
   ISSN={1538-4357},
   url={http://dx.doi.org/10.3847/1538-4357/ac17ea},
   DOI={10.3847/1538-4357/ac17ea},
   number={1},
   journal={The Astrophysical Journal},
   publisher={American Astronomical Society},
   author={Abbott, R. and others },
   year={2021},
   month=nov, pages={80} }

@misc{AEIULurl,
author = {J.~Ming and others},
year           = "2025",
HOWPUBLISHED = "\url{www.aei.mpg.de/continuouswaves/O3aCasaVelajrG3473}"

}

@article{Abbott2022_4,
  title = "{Search of the early O3 LIGO data for continuous gravitational waves from the Cassiopeia A and Vela Jr. supernova remnants}",
  author = {Abbott, R. and others},
  collaboration = {LIGO Scientific Collaboration and Virgo Collaboration},
  journal = {Phys. Rev. D},
  volume = {105},
  issue = {8},
  pages = {082005},
  numpages = {25},
  year = {2022},
  month = {Apr},
  publisher = {American Physical Society},
  doi = {10.1103/PhysRevD.105.082005},
  url = {https://link.aps.org/doi/10.1103/PhysRevD.105.082005}
}

@article{Ming2019,
  title = {Results from an Einstein@Home search for continuous gravitational waves from Cassiopeia A, Vela Jr., and G347.3},
  author = {Ming, J. and Papa, M. A. and Singh, A. and Eggenstein, H.-B. and Zhu, S. J. and Dergachev, V. and Hu, Y. and Prix, R. and Machenschalk, B. and Beer, C. and Behnke, O. and Allen, B.},
  journal = {Phys. Rev. D},
  volume = {100},
  issue = {2},
  pages = {024063},
  numpages = {11},
  year = {2019},
  month = {Jul},
  publisher = {American Physical Society},
  doi = {10.1103/PhysRevD.100.024063},
  url = {https://link.aps.org/doi/10.1103/PhysRevD.100.024063}
}

@article{Ming2022,
doi = {10.3847/1538-4357/ac35cb},
url = {https://dx.doi.org/10.3847/1538-4357/ac35cb},
year = {2022},
month = {jan},
publisher = {The American Astronomical Society},
volume = {925},
number = {1},
pages = {8},
author = {J. Ming and M. A. Papa and H.-B. Eggenstein and B. Machenschalk and B. Steltner and R. Prix and B. Allen and O. Behnke},
title = "{Results From an Einstein@Home Search for Continuous Gravitational Waves From G347.3 at Low Frequencies in LIGO O2 Data}",
journal = {The Astrophysical Journal},
}

@ARTICLE{Ushomirsky2000,
   author       = "G. Ushomirsky and C. Cutler and L. Bildsten", 
   journal      = "Mon. Notices Royal Astron. Soc.",
   title        = "{Deformations of Accreting Neutron Star Crusts and Gravitational Wave Emission}", 
   year         = "2000",

   volume       = "319", 
   pages        = "902--932",
   month        = {Aug},
   doi = {10.1046/j.1365-8711.2000.03938.x},
}

@article{Haskell2006,
    author = {Haskell, B. and Jones, D. I. and Andersson, N.},
    title = "{Mountains on neutron stars: accreted versus non-accreted crusts}",
    journal = {Monthly Notices of the Royal Astronomical Society},
    volume = {373},
    number = {4},
    pages = {1423-1439},
    year = {2006},
    month = {11},
    issn = {0035-8711},
    doi = {10.1111/j.1365-2966.2006.10998.x},
    url = {https://doi.org/10.1111/j.1365-2966.2006.10998.x},
    eprint = {https://academic.oup.com/mnras/article-pdf/373/4/1423/11177018/mnras0373-1423.pdf},
}

@article{Arras_2003,
doi = {10.1086/374657},
url = {https://dx.doi.org/10.1086/374657},
year = {2003},
month = {jul},
publisher = {},
volume = {591},
number = {2},
pages = {1129},
author = {Phil Arras and Eanna E. Flanagan and Sharon M. Morsink and A. Katrin Schenk and Saul A. Teukolsky and Ira Wasserman},
title = {Saturation of the r-Mode Instability},
journal = {The Astrophysical Journal}
}

@misc{Owen:1998xg,
      author         = "Owen, Benjamin J. and Lindblom, Lee and Cutler, Curt and
                        Schutz, Bernard F. and Vecchio, Alberto and Andersson,
                        Nils",
      title          = "{Gravitational waves from hot young rapidly rotating
                        neutron stars}",
      journal        = "Phys. Rev.",
      volume         = "D58",
      year           = "1998",
      pages          = "084020",
      doi            = "10.1103/PhysRevD.58.084020",
      eprint         = "gr-qc/9804044",
      archivePrefix  = "arXiv",
      primaryClass   = "gr-qc",
      reportNumber   = "GRP-496",
      SLACcitation   = "%%CITATION = GR-QC/9804044;%%"
}

@article{PhysRevD.76.064019,
  title = {Spin evolution of accreting neutron stars: Nonlinear development of the $r$-mode instability},
  author = {Bondarescu, Ruxandra and Teukolsky, Saul A. and Wasserman, Ira},
  journal = {Phys. Rev. D},
  volume = {76},
  issue = {6},
  pages = {064019},
  numpages = {18},
  year = {2007},
  month = {Sep},
  publisher = {American Physical Society},
  doi = {10.1103/PhysRevD.76.064019},
  url = {https://link.aps.org/doi/10.1103/PhysRevD.76.064019}
}

@article{Haskell:2015iia,
    author = "Haskell, B.",
    title = "{R-modes in neutron stars: Theory and observations}",
    eprint = "1509.04370",
    archivePrefix = "arXiv",
    primaryClass = "astro-ph.HE",
    doi = "10.1142/S0218301315410074",
    journal = "Int. J. Mod. Phys. E",
    volume = "24",
    number = "09",
    pages = "1541007",
    year = "2015"
}

@article{Morales2022,
    author = {Morales, J A and Horowitz, C J},
    title = "{Neutron star crust can support a large ellipticity}",
    journal = {Monthly Notices of the Royal Astronomical Society},
    volume = {517},
    number = {4},
    pages = {5610-5616},
    year = {2022},
    month = {10},
    issn = {0035-8711},
    doi = {10.1093/mnras/stac3058},
    url = {https://doi.org/10.1093/mnras/stac3058},
    eprint = {https://academic.oup.com/mnras/article-pdf/517/4/5610/46951663/stac3058.pdf},
}

@article{Morales2024,
  title = {Anisotropic neutron star crust, solar system mountains, and gravitational waves},
  author = {Morales, J. A. and Horowitz, C. J.},
  journal = {Phys. Rev. D},
  volume = {110},
  issue = {4},
  pages = {044016},
  numpages = {8},
  year = {2024},
  month = {Aug},
  publisher = {American Physical Society},
  doi = {10.1103/PhysRevD.110.044016},
  url = {https://link.aps.org/doi/10.1103/PhysRevD.110.044016}
}

@article{Andersson1999,
doi = {10.1086/307082},
url = {https://dx.doi.org/10.1086/307082},
year = {1999},
month = {may},
publisher = {},
volume = {516},
number = {1},
pages = {307},
author = {Nils Andersson and Kostas D. Kokkotas and Nikolaos Stergioulas},
title = {On the Relevance of the r-Mode Instability for Accreting Neutron Stars and White Dwarfs},
journal = {The Astrophysical Journal}
}

@article{Owen2010,
  title = "{How to adapt broad-band gravitational-wave searches for $r$-modes}",
  author = {Owen, Benjamin J.},
  journal = {Phys. Rev. D},
  volume = {82},
  issue = {10},
  pages = {104002},
  numpages = {8},
  year = {2010},
  month = {Nov},
  publisher = {American Physical Society},
  doi = {10.1103/PhysRevD.82.104002},
  url = {https://link.aps.org/doi/10.1103/PhysRevD.82.104002}
}

@article{Gittins2023,
    author = {Gittins, Fabian and Andersson, Nils},
    title = "{The r-modes of slowly rotating, stratified neutron stars}",
    journal = {Monthly Notices of the Royal Astronomical Society},
    volume = {521},
    number = {2},
    pages = {3043-3057},
    year = {2023},
    month = {03},
    issn = {0035-8711},
    doi = {10.1093/mnras/stad672},
    url = {https://doi.org/10.1093/mnras/stad672},
    eprint = {https://academic.oup.com/mnras/article-pdf/521/2/3043/49597494/stad672.pdf},
}

@ARTICLE{ming2025,
       author = {{Morales}, Jorge and {Ming}, Jing and {Papa}, Maria Alessandra and {Eggenstein}, Heinz-Bernd and {Machenschalk}, Bernd},
        title = "{Results from an Einstein@Home Search for Continuous Gravitational Waves from Cassiopeia A and Vela Jr. Using LIGO O2 Data}",
      journal = {\apj},
         year = 2025,
        month = jun,
       volume = {986},
       number = {2},
          eid = {202},
        pages = {202},
          doi = {10.3847/1538-4357/add5f5},
archivePrefix = {arXiv},
       eprint = {2503.09731},
 primaryClass = {gr-qc},
}

@article{Beheshtipour:2020zhb,
    author = "Beheshtipour, Banafsheh and Papa, Maria Alessandra",
    title = "{Deep learning for clustering of continuous gravitational wave candidates}",
    eprint = "2001.03116",
    archivePrefix = "arXiv",
    primaryClass = "gr-qc",
    doi = "10.1103/PhysRevD.101.064009",
    journal = "Phys. Rev. D",
    volume = "101",
    number = "6",
    pages = "064009",
    year = "2020"
}

@article{Beheshtipour:2020nko,
    author = "Beheshtipour, B. and Papa, M. A.",
    title = "{Deep learning for clustering of continuous gravitational wave candidates II: identification of low-SNR candidates}",
    eprint = "2012.04381",
    archivePrefix = "arXiv",
    primaryClass = "gr-qc",
    doi = "10.1103/PhysRevD.103.064027",
    journal = "Phys. Rev. D",
    volume = "103",
    number = "6",
    pages = "064027",
    year = "2021"
}

@article{Ming2016,
  title = {Optimal directed searches for continuous gravitational waves},
  author = {Ming, Jing and Krishnan, Badri and Papa, Maria Alessandra and Aulbert, Carsten and Fehrmann, Henning},
  journal = {Phys. Rev. D},
  volume = {93},
  issue = {6},
  pages = {064011},
  numpages = {26},
  year = {2016},
  month = {Mar},
  publisher = {American Physical Society},
  doi = {10.1103/PhysRevD.93.064011},
  url = {https://link.aps.org/doi/10.1103/PhysRevD.93.064011}
}

@article{Steltner2022,
  title = {Identification and removal of non-Gaussian noise transients for gravitational-wave searches},
  author = {Steltner, Benjamin and Papa, Maria Alessandra and Eggenstein, Heinz-Bernd},
  journal = {Phys. Rev. D},
  volume = {105},
  issue = {2},
  pages = {022005},
  numpages = {10},
  year = {2022},
  month = {Jan},
  publisher = {American Physical Society},
  doi = {10.1103/PhysRevD.105.022005},
  url = {https://link.aps.org/doi/10.1103/PhysRevD.105.022005}
}

@article{Pletsch2008,
  title = {Parameter-space correlations of the optimal statistic for continuous gravitational-wave detection},
  author = {Pletsch, Holger J.},
  journal = {Phys. Rev. D},
  volume = {78},
  issue = {10},
  pages = {102005},
  numpages = {17},
  year = {2008},
  month = {Nov},
  publisher = {American Physical Society},
  doi = {10.1103/PhysRevD.78.102005},
  url = {https://link.aps.org/doi/10.1103/PhysRevD.78.102005}
}

@article{Pletsch2009,
  title = {Exploiting Large-Scale Correlations to Detect Continuous Gravitational Waves},
  author = {Pletsch, Holger J. and Allen, Bruce},
  journal = {Phys. Rev. Lett.},
  volume = {103},
  issue = {18},
  pages = {181102},
  numpages = {4},
  year = {2009},
  month = {Oct},
  publisher = {American Physical Society},
  doi = {10.1103/PhysRevLett.103.181102},
  url = {https://link.aps.org/doi/10.1103/PhysRevLett.103.181102}
}

@article{Pletsch2010,
  title = {Parameter-space metric of semicoherent searches for continuous gravitational waves},
  author = {Pletsch, Holger J.},
  journal = {Phys. Rev. D},
  volume = {82},
  issue = {4},
  pages = {042002},
  numpages = {14},
  year = {2010},
  month = {Aug},
  publisher = {American Physical Society},
  doi = {10.1103/PhysRevD.82.042002},
  url = {https://link.aps.org/doi/10.1103/PhysRevD.82.042002}
}

@article{JKS1998,
  title = {Data analysis of gravitational-wave signals from spinning neutron stars: The signal and its detection},
  author = {Jaranowski, Piotr and Kr\'olak, Andrzej and Schutz, Bernard F.},
  journal = {Phys. Rev. D},
  volume = {58},
  issue = {6},
  pages = {063001},
  numpages = {24},
  year = {1998},
  month = {Aug},
  publisher = {American Physical Society},
  doi = {10.1103/PhysRevD.58.063001},
  url = {https://link.aps.org/doi/10.1103/PhysRevD.58.063001}
}

@INPROCEEDINGS{Anderson2004,
  author={Anderson, D.P.},
  booktitle={Fifth IEEE/ACM International Workshop on Grid Computing}, 
  title={BOINC: a system for public-resource computing and storage}, 
  year={2004},
  volume={},
  number={},
  pages={4-10},
  doi={10.1109/GRID.2004.14}}

@inproceedings{Anderson2006,
author = {Anderson, David P. and Christensen, Carl and Allen, Bruce},
title = {Designing a runtime system for volunteer computing},
year = {2006},
isbn = {0769527000},
publisher = {Association for Computing Machinery},
address = {New York, NY, USA},
url = {https://doi.org/10.1145/1188455.1188586},
doi = {10.1145/1188455.1188586},
booktitle = {Proceedings of the 2006 ACM/IEEE Conference on Supercomputing},
pages = {126–es},
location = {Tampa, Florida},
series = {SC '06}
}

@ARTICLE{Tanabaum1999,
       author = {{Tananbaum}, H.},
        title = "{Cassiopeia A}",
      journal = {IAU Cirulars},
         year = 1999,
        month = sep,
       volume = {7246},
        pages = {1},
       adsurl = {https://ui.adsabs.harvard.edu/abs/1999IAUC.7246....1T}
}

@article{Fesen2006,
doi = {10.1086/504254},
url = {https://dx.doi.org/10.1086/504254},
year = {2006},
month = {jul},
publisher = {},
volume = {645},
number = {1},
pages = {283},
author = {Robert A. Fesen and Molly C. Hammell and Jon Morse and Roger A. Chevalier and Kazimierz J. Borkowski and Michael A. Dopita and Christopher L. Gerardy and Stephen S. Lawrence and John C. Raymond and Sidney van den Bergh},
title = {The Expansion Asymmetry and Age of the Cassiopeia A Supernova Remnant*},
journal = {The Astrophysical Journal}
}

@ARTICLE{Reed1995,
       author = {{Reed}, Jeri E. and {Hester}, J. Jeff and {Fabian}, A.~C. and {Winkler}, P.~F.},
        title = "{The Three-dimensional Structure of the Cassiopeia A Supernova Remnant. I. The Spherical Shell}",
      journal = {\apj},
         year = 1995,
        month = feb,
       volume = {440},
        pages = {706},
          doi = {10.1086/175308},
       adsurl = {https://ui.adsabs.harvard.edu/abs/1995ApJ...440..706R}
}

@article{Ho2009,
  title={A neutron star with a carbon atmosphere in the Cassiopeia A supernova remnant},
  author={Ho, Wynn and Heinke, Craig},
  journal={Nature},
  volume={462},
  pages={71--73},
  year={2009},
  publisher={Nature Publishing Group},
  doi={10.1038/nature08525},
  url={https://doi.org/10.1038/nature08525},
  received={21 July 2009},
  accepted={15 September 2009},
  issue_date={05 November 2009}
}

@article{Pavlov2001,
doi = {10.1086/323975},
url = {https://dx.doi.org/10.1086/323975},
year = {2001},
month = {sep},
publisher = {},
volume = {559},
number = {2},
pages = {L131},
author = {George G. Pavlov and Divas Sanwal and Bülent Kızıltan and Gordon P. Garmire},
title = {The Compact Central Object in the RX J0852.0–4622 Supernova Remnant},
journal = {The Astrophysical Journal}
}

@article{Iyudin1998,
  author={A. F. Iyudin and V. Schoenfelder and K. Bennett and H. Bloemen and R. Diehl and W. Hermsen and G. G. Lichti and R. D. van der Meulen and J. Ryan and C. Winkler},
  title={{Emission from 44Ti associated with a previously unknown Galactic supernova}},
  journal={Nature},
  year=1998,
  volume={396},
  number={6707},
  pages={142-144},
  month={November},
  doi={10.1038/24106},
}

@article{Allen2015,
   title={ON THE EXPANSION RATE, AGE, AND DISTANCE OF THE SUPERNOVA REMNANT G266.2?1.2 (Vela Jr.)},
   volume={798},
   ISSN={1538-4357},
   url={http://dx.doi.org/10.1088/0004-637X/798/2/82},
   DOI={10.1088/0004-637x/798/2/82},
   number={2},
   journal={The Astrophysical Journal},
   publisher={American Astronomical Society},
   author={Allen, G. E. and Chow, K. and DeLaney, T. and Filipovi?, M. D. and Houck, J. C. and Pannuti, T. G. and Stage, M. D.},
   year={2014},
   month=dec, pages={82} }

@article{Ashok2024,
  title = {Bayesian $\mathcal{F}$-statistic-based parameter estimation of continuous gravitational waves from known pulsars},
  author = {Ashok, A. and Covas, P. B. and Prix, R. and Papa, M. A.},
  journal = {Phys. Rev. D},
  volume = {109},
  issue = {10},
  pages = {104002},
  numpages = {15},
  year = {2024},
  month = {May},
  publisher = {American Physical Society},
  doi = {10.1103/PhysRevD.109.104002},
  url = {https://link.aps.org/doi/10.1103/PhysRevD.109.104002}
}

@article{Mignani2007,
	author = {{Mignani, R. P.} and {De Luca, A.} and {Zaggia, S.} and {Sester, D.} and {Pellizzoni, A.} and {Mereghetti, S.} and {Caraveo, P. A.}},
	title = {VLT observations of the central compact object  in the Vela Jr. supernova remnant *},
	DOI= "10.1051/0004-6361:20077768",
	url= "https://doi.org/10.1051/0004-6361:20077768",
	journal = {Astronomy and Astrophysics},
	year = 2007,
	volume = 473,
	number = 3,
	pages = "883-889",
}

@misc{E@H,
  author  = "Einstein@Home",
  title = {Einstein@Home},
  howpublished = {\url{https://einsteinathome.org/}},
  year           = "2023",
}

@misc{Atlas,
	author  = "Allen, B.",
	HOWPUBLISHED = "\url{https://www.aei.mpg.de/25950/computer-clusters}",
	TITLE = "Computing and ATLAS",
	year           = "2025",
}

@article{Behnke:2014tma,
    author = "Behnke, Berit and Papa, Maria Alessandra and Prix, Reinhard",
    title = "{Postprocessing methods used in the search for continuous gravitational-wave signals from the Galactic Center}",
    eprint = "1410.5997",
    archivePrefix = "arXiv",
    primaryClass = "gr-qc",
    reportNumber = "LIGO-DOCUMENT-NUMBER-P1300125, AEI-2014-017",
    doi = "10.1103/PhysRevD.91.064007",
    journal = "Phys. Rev. D",
    volume = "91",
    number = "6",
    pages = "064007",
    year = "2015"
}

@article{Brady:1997ji,
    author = "Brady, Patrick R. and Creighton, Teviet and Cutler, Curt and Schutz, Bernard F.",
    title = "{Searching for periodic sources with LIGO}",
    eprint = "gr-qc/9702050",
    archivePrefix = "arXiv",
    reportNumber = "GRP-460, CGPG-97-3-2",
    doi = "10.1103/PhysRevD.57.2101",
    journal = "Phys. Rev. D",
    volume = "57",
    pages = "2101--2116",
    year = "1998"
}

@article{Cutler:2005hc,
      author         = "Cutler, Curt and Schutz, Bernard F.",
      title          = "{The Generalized F-statistic: Multiple detectors and
                        multiple GW pulsars}",
      journal        = "Phys. Rev.",
      volume         = "D72",
      year           = "2005",
      pages          = "063006",
      doi            = "10.1103/PhysRevD.72.063006",
      eprint         = "0504011",
      archivePrefix  = "arXiv",
      primaryClass   = "gr-qc",
      SLACcitation   = "%%CITATION = GR-QC/0504011;%%"
}

@ARTICLE{2000PhRvD..61h2001B,
       author = {{Brady}, Patrick R. and {Creighton}, Teviet},
        title = "{Searching for periodic sources with LIGO. II. Hierarchical searches}",
      journal = {\prd},
         year = 2000,
        month = apr,
       volume = {61},
       number = {8},
          eid = {082001},
        pages = {082001},
          doi = {10.1103/PhysRevD.61.082001},
archivePrefix = {arXiv},
       eprint = {gr-qc/9812014},
 primaryClass = {gr-qc},
       adsurl = {https://ui.adsabs.harvard.edu/abs/2000PhRvD..61h2001B}
}

@article{Dreissigacker2018,
  title = {Fast and accurate sensitivity estimation for continuous-gravitational-wave searches},
  author = {Dreissigacker, Christoph and Prix, Reinhard and Wette, Karl},
  journal = {Phys. Rev. D},
  volume = {98},
  issue = {8},
  pages = {084058},
  numpages = {30},
  year = {2018},
  month = {Oct},
  publisher = {American Physical Society},
  doi = {10.1103/PhysRevD.98.084058},
  url = {https://link.aps.org/doi/10.1103/PhysRevD.98.084058}
}

@misc{Keitel:2013wga,
      author         = "Keitel, David and Prix, Reinhard and Papa, Maria
                        Alessandra and Leaci, Paola and Siddiqi, Maham",
      title          = "{Search for continuous gravitational waves: Improving
                        robustness versus instrumental artifacts}",
      journal        = "Phys. Rev.",
      volume         = "D89",
      year           = "2014",
      number         = "6",
      pages          = "064023",
      doi            = "10.1103/PhysRevD.89.064023",
      eprint         = "1311.5738",
      archivePrefix  = "arXiv",
      primaryClass   = "gr-qc",
      reportNumber   = "LIGO-P1300167, AEI-2013-260",
      SLACcitation   = "%%CITATION = ARXIV:1311.5738;%%"
}

@article{Keitel:2015ova,
      author         = "Keitel, David",
      title          = "{Robust semicoherent searches for continuous
                        gravitational waves with noise and signal models including
                        hours to days long transients}",
      journal        = "Phys. Rev.",
      volume         = "D93",
      year           = "2016",
      number         = "8",
      pages          = "084024",
      doi            = "10.1103/PhysRevD.93.084024",
      eprint         = "1509.02398",
      archivePrefix  = "arXiv",
      primaryClass   = "gr-qc",
      reportNumber   = "LIGO-P1500159",
      SLACcitation   = "%%CITATION = ARXIV:1509.02398;%%"
}

@misc{o3_linefile,
	author  = "LVK",
	HOWPUBLISHED = "\url{https://gwosc.org/O3/o3speclines/}",
	TITLE = "O3 Instrumental Lines",
	year           = "2021",
}

@misc{o4_linefile,
	author  = "LVK",
	HOWPUBLISHED = "\url{https://gwosc.org/O4/o4speclines/}",
	TITLE = "O4 Instrumental Lines",
	year           = "2025",
}

@article{Vajente_2020,
  title = {Machine-learning nonstationary noise out of gravitational-wave detectors},
  author = {Vajente, G. and Huang, Y. and Isi, M. and Driggers, J. C. and Kissel, J. S. and Szczepa\ifmmode \acute{n}\else \'{n}\fi{}czyk, M. J. and Vitale, S.},
  journal = {Phys. Rev. D},
  volume = {101},
  issue = {4},
  pages = {042003},
  numpages = {12},
  year = {2020},
  month = {Feb},
  publisher = {American Physical Society},
  doi = {10.1103/PhysRevD.101.042003},
  url = {https://link.aps.org/doi/10.1103/PhysRevD.101.042003}
}

@article{Davis_2019,
	doi = {10.1088/1361-6382/ab01c5},
	url = {https://doi.org/10.1088/1361-6382/ab01c5},
	year = 2019,
	month = {feb},
	publisher = {{IOP} Publishing},
	volume = {36},
	number = {5},
	pages = {055011},
	author = {Derek Davis and Thomas Massinger and Andrew Lundgren and Jennifer C Driggers and Alex L Urban and Laura Nuttall},
	title = {Improving the sensitivity of Advanced {LIGO} using noise subtraction},
	journal = {Classical and Quantum Gravity}
}

@article{DiCesare:2025wnb,
    author = "Di Cesare, Martina",
    title = "{Status of the O4 run and latest non-CBC results}",
    eprint = "2505.18802",
    archivePrefix = "arXiv",
    primaryClass = "gr-qc",
    month = "5",
    year = "2025"
}

@article{LIGOScientific:2025pvj,
    author = "Abac, A. G. and others",
    collaboration = "LIGO Scientific, VIRGO, KAGRA",
    title = "{GWTC-4.0: Population Properties of Merging Compact Binaries}",
    eprint = "2508.18083",
    archivePrefix = "arXiv",
    primaryClass = "astro-ph.HE",
    reportNumber = "LIGO-P2400004",
    month = "8",
    year = "2025"
}

@ARTICLE{2023PhRvX..13d1039A,
       author = {Abbott, Rich and others},
        title = "{GWTC-3: Compact Binary Coalescences Observed by LIGO and Virgo during the Second Part of the Third Observing Run}",
      journal = {Physical Review X},
         year = 2023,
        month = oct,
       volume = {13},
       number = {4},
          eid = {041039},
        pages = {041039},
          doi = {10.1103/PhysRevX.13.041039},
archivePrefix = {arXiv},
       eprint = {2111.03606},
 primaryClass = {gr-qc},
       adsurl = {https://ui.adsabs.harvard.edu/abs/2023PhRvX..13d1039A}
}

@ARTICLE{ming2024a,
       author = {{Ming}, J. and {Papa}, M.~A. and {Eggenstein}, H. -B. and {Beheshtipour}, B. and {Machenschalk}, B. and {Prix}, R. and {Allen}, B. and {Bensch}, M.},
        title = "{Deep Einstein@Home Search for Continuous Gravitational Waves from the Central Compact Objects in the Supernova Remnants Vela Jr. and G347.3-0.5 Using LIGO Public Data}",
      journal = {\apj},
         year = 2024,
        month = dec,
       volume = {977},
       number = {2},
          eid = {154},
        pages = {154},
          doi = {10.3847/1538-4357/ad8b9e},
archivePrefix = {arXiv},
       eprint = {2408.14573},
 primaryClass = {gr-qc},
       adsurl = {https://ui.adsabs.harvard.edu/abs/2024ApJ...977..154M}
}

@article{Singh:2017kss,
    author = "Singh, Avneet and Papa, Maria Alessandra and Eggenstein, Heinz-Bernd and Walsh, Sin\'ead",
    title = "{Adaptive clustering procedure for continuous gravitational wave searches}",
    eprint = "1707.02676",
    archivePrefix = "arXiv",
    primaryClass = "gr-qc",
    reportNumber = "LIGO-DOCUMENT-LIGO-P1700123",
    doi = "10.1103/PhysRevD.96.082003",
    journal = "Phys. Rev. D",
    volume = "96",
    number = "8",
    pages = "082003",
    year = "2017"
}

@article{den_cluster,
  title = {Density-clustering of continuous gravitational wave candidates from large surveys},
  author = {Steltner, B. and Menne, T. and Papa, M. A. and Eggenstein, H.-B.},
  journal = {Phys. Rev. D},
  volume = {106},
  issue = {10},
  pages = {104063},
  numpages = {8},
  year = {2022},
  month = {Nov},
  publisher = {American Physical Society},
  doi = {10.1103/PhysRevD.106.104063},
  url = {https://link.aps.org/doi/10.1103/PhysRevD.106.104063}
}

@article{Papa_2020midth,
	doi = {10.3847/1538-4357/ab92a6},
	url = {https://doi.org/10.3847/1538-4357/ab92a6},
	year = 2020,
	month = {jun},
	publisher = {American Astronomical Society},
	volume = {897},
	number = {1},
	pages = {22},
	author = {M. A. Papa and J. Ming and E. V. Gotthelf and B. Allen and R. Prix and V. Dergachev and H.-B. Eggenstein and A. Singh and S. J. Zhu},
	title = {Search for Continuous Gravitational Waves from the Central Compact Objects in Supernova Remnants Cassiopeia A, Vela Jr., and G347.3{\textendash}0.5},
	journal = {The Astrophysical Journal}
}

@article{Vargas:2024itq,
    author = "Vargas, Andr\'es F. and Melatos, Andrew",
    title = "{Stochastic and secular anomalies in pulsar braking indices}",
    eprint = "2410.04757",
    archivePrefix = "arXiv",
    primaryClass = "astro-ph.HE",
    doi = "10.1093/mnras/stae2326",
    journal = "Mon. Not. Roy. Astron. Soc.",
    volume = "534",
    number = "4",
    pages = "3410--3422",
    year = "2024"
}
\bibliographystyle{aasjournal}

\end{document}